\documentclass[aps,nofootinbib,physrev,twocolumn,groupedaddress]{revtex4-2}
\usepackage{amsmath}
\usepackage{standalone}
\usepackage{currfile}
\usepackage{tikz}
\usetikzlibrary{math,calc,arrows.meta,decorations.text,3d,perspective,shadings,decorations.markings,bending}
\definecolor{CGreen}{RGB}{0,100,0}
\usepackage{pgfplots}
\pgfplotsset{compat=1.16}
\usepackage{graphicx}
\usepackage{dcolumn}  
\usepackage{bm}  
\usepackage{lipsum}
\usepackage{import}
\usepackage{siunitx}
\usepackage{hyperref}
\usepackage{lipsum}
\usepackage{float}
\usepackage{xfrac}
\hypersetup{colorlinks=true, citecolor=blue, urlcolor=blue, linkcolor=blue}
\usepackage[shortlabels]{enumitem}
\usepackage{braket}
\usepackage{scalefnt}
\usepackage{ifthen}
\usepackage{calligra}
\usepackage[T1]{fontenc}
\usepackage{frcursive}

\usepackage{mathtools}
\usepackage{dsfont}
\usepackage{substr}

\newcommand{\vect}[1]{\vec{\boldsymbol{\mathbf{#1}}}}
\newcommand{\uvect}[1]{\hat{\boldsymbol{\mathbf{#1}}}}

\usepackage{amssymb}

\makeatletter
\newcommand\makesymbolaccent[2]{%
  \expandafter\expandafter\expandafter\let
  \expandafter\expandafter\expandafter\@tempa\expandafter\@secondoftwo#2%
  \edef#1{{\mkern-1mu\noexpand\symbolaccent{\mathchar"\expandafter\@gobblethree\@tempa}}}%
}
\newcommand{\symbolaccent}[1]{\mathpalette\symbol@accent{#1}}
\newcommand{\symbol@accent}[2]{%
  \sbox\z@{$\m@th\demote@style{#1}#2$}%
  \sbox\tw@{$\m@th\demote@style{#1}{}^{\demote@style{#1}#2}$}%
  \raisebox{\dimexpr\ht\z@-\ht\tw@}[\height][0pt]{\copy\z@}%
}
\newcommand\demote@style[1]{%
  \ifx#1\scriptstyle\textstyle\else\scriptstyle\fi
}
\makeatother

\makesymbolaccent{\ehat}{\hat}
\makesymbolaccent{\etilde}{\tilde}
\makesymbolaccent{\ebreve}{\breve}
\makesymbolaccent{\egrave}{\grave}

\DeclareMathOperator{\sgn}{sgn}

\usepackage{jobname-suffix}
\newcommand{\markstartchange}{\color{black}}
\newcommand{\markstopchange}{\color{black}}

\IfSuffixT[highlight_generated]{
  \renewcommand{\markstartchange}{\color{blue}}
  \renewcommand{\markstopchange}{\color{black}}
}

\begin{document}

\title{Stern-Gerlach interferometry in three dimensions: the role of transverse fields}

\author{D.~Meng}
\thanks{\href{mailto:d22meng@uwaterloo.ca}{d22meng@uwaterloo.ca}}
\affiliation{Department of Physics and Astronomy, University of Waterloo, Waterloo N2L 3G1 Canada}
\author{D.~Z.~Chan}
\thanks{Present address: \emph{Department of Physics, University of Oxford, Denys Wilkinson Building, Keble Road, Oxford OX1 3RH, United Kingdom}, \href{mailto:darren.chan@physics.ox.ac.uk}{darren.chan@physics.ox.ac.uk} and \href{mailto:dzchan@uwaterloo.ca}{dzchan@uwaterloo.ca}}
\author{J.~D.~D.~Martin}
\thanks{\href{mailto:jddmartin@uwaterloo.ca}{jddmartin@uwaterloo.ca}}
\affiliation{Department of Physics and Astronomy, University of Waterloo, Waterloo N2L 3G1 Canada}

\date{\today}

\begin{abstract}
We show that superficially similar implementations of Stern-Gerlach Interferometers (SGIs) are expected to differ dramatically in their sensitivity to fields transverse to the primary acceleration direction.  These transverse fields unavoidably accompany any static magnetic or electric field gradients, and have been shown by Comparat [\href{https://link.aps.org/doi/10.1103/PhysRevA.101.023606}{Phys.~Rev.~A \textbf{101}, 023606 (2020)}] to limit the precision application of SGIs. As a concrete example, we consider SGIs with ultracold Rb Rydberg atoms accelerated by spatially-varying electric fields. We find that the deleterious effect of transverse fields imply that only some implementations (sequences of field gradients, internal state swaps, and so-on) may exhibit fringes with high visibility.

\end{abstract}

\maketitle

\section{Introduction}

Although Humpty-Dumpty is commonly portrayed as an egg, the original nursery rhyme is a riddle with his identity unspecified \cite{isbn:9780198691129}.  Following Englert {\it et al.}~\cite{shortdoi:c34q7x}, we like to think of him as a state-vector describing both the position and internal state of an atom.  And instead of falling off a wall, he suffers the equally calamitous fate of Stern-Gerlach splitting.

\begin{figure}
\includegraphics{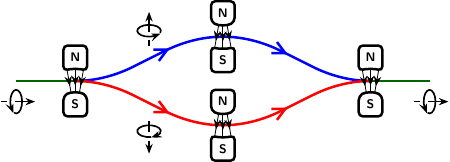}
\caption{\label{fg:sg_apparatus} An SGI as envisioned by early workers illustrating the splitting and recombination of two spin states \cite{isbn:9780486659695,shortdoi:bdgdmp,shortdoi:c34q7x}.}
\end{figure}

As illustrated in Fig.~\ref{fg:sg_apparatus}, ``putting Humpty-Dumpty back together again'' corresponds to the construction of a so-called Stern-Gerlach interferometer (SGI).  The desired final state is a coherent superposition of $\ket{\uparrow}$ and $\ket{\downarrow}$ states, each with the same spatial wavefunction.  The difficulty in such a reconstruction was first studied in detail by Englert {\it et al.}~\cite{shortdoi:c34q7x}, who concluded that it may be possible, provided a ``fantastic precision of one part in \SI{e5}{} for $\partial B / \partial z$''.  Unsurprisingly, there has been no experimental implementation of an SGI in the form shown in Fig.~\ref{fg:sg_apparatus}.

However, the principle of Galilean relativity suggests an equivalent but more realizable experiment: atoms with no net motion are exposed to a sequence of spatially-varying fields, obtained by the precise control of currents through coils, mimicking the time-dependence caused by flight through the different magnets in Fig.~\ref{fg:sg_apparatus}.  There is no longer a need for stringent magnetic field control at different locations along a beam.  SGIs in this form have been recently realized using ultra-cold atoms and magnetic micro-traps \cite{shortdoi:gn7jtg, shortdoi:gkcvdb}.

In contrast to SGIs, matter-wave interferometers using photon-recoil based beam-splitters (and combiners) have been extensively developed and have achieved exceptional performance, particularly for measurements of the acceleration due to gravity $g$ \cite{shortdoi:fdtkbw} (for recent progress, see Ref.~\cite{shortdoi:ggxx4v} and references therein).  A key parameter dictating the sensitivity of these matter-wave interferometers to $g$ is the interferometric ``area'' between their two paths.


It has been suggested \cite{shortdoi:gpb7cs} that SGIs with continuous accelerations may obtain larger interferometric areas and thus even higher sensitivities to $g$ than photon-recoil based interferometers. To this end, Comparat \cite{shortdoi:gptnmb} studied an SGI that would use spatially-varying \emph{electric} fields to accelerate H Rydberg atoms with large electric dipole moments ($10^2$ to $10^3$ $q a_0$ where $q$ is the elementary charge and $a_0$ is the Bohr radius). These strong accelerations result in the large interferometric area of the SGI. But Comparat predicted that this particular SGI has only mildly better sensitivity to $g$ than certain photon-recoil based interferometers --- with some limitations to SGI arising because of accelerations by fields in directions transverse to the accelerating field gradient. These fields are an unavoidable consequence of the equations of electrostatics. For example, in a cylindrically symmetric geometry, $\vect{\nabla} \cdot \vect{E} = 0$ implies that along the symmetry axis $z$ we have $\partial_x E_x = \partial_y E_y = -(1/2) \partial_z E_z$. Thus, given a non-zero $\partial_z E_z$ required for acceleration, once we are off the symmetry axis, we have non-zero transverse field components $E_x$ and $E_y$.

In this paper, we show that choices in the implementation of SGIs can dramatically influence the deleterious effect of these transverse fields on the interferometric visibility.  We derive general closed form expressions for fringe visibilities and make our results concrete by analyzing a proposed SGI using Rb Rydberg atoms obtained by excitation of translationally cold atoms from a magneto-optical trap (MOT).  Most of our results are also applicable to magnetic field SGIs \cite{shortdoi:gn7jtg, shortdoi:gkcvdb, shortdoi:gt4jhh}
due to the similarity of the equations of magneto- and electrostatics.

It is desirable to establish the universality of the limitations on SGIs due to transverse fields, especially since SGIs could be implemented in significantly different physical systems from those initially demonstrated in Refs~\cite{shortdoi:gn7jtg, shortdoi:gkcvdb} and proposed in Ref.~\cite{shortdoi:gptnmb}.  For example, pairs of SGIs have been proposed to test the quantum nature of gravity \cite{shortdoi:gcsb22}.  These SGIs will accelerate nanodiamonds using inhomogeneous magnetic fields acting on embedded spins associated with nitrogen-vacancy (NV) centers (see Bose \emph{et al.}~\cite{shortdoi:g94nnc_alt} and references therein for recent progress). \markstartchange Other configurations have been proposed for precision field gradient measurements \cite{shortdoi:g9hccm} and direct detection of axion-like dark matter \cite{shortdoi:hbdsg9}. \markstopchange

\markstartchange
Many challenges in realizing SGIs have been identified, with additional loss of coherence stemming from sources including internal phonons \cite{xiang_phonon_induced_2024, henkel_internal_2022, henkel_universal_2024}, noise from the experiment apparatus itself \cite{wu_inertial_2025, henkel_fundamental_2003, narasimha_moorthy_magnetic_2026, zhou_spin_2025}, and rotational degrees of freedom \cite{zhou_gyroscopic_2025, japha_quantum_2023, rizaldy_rotational_2025}, among others. Analysis of the problem of coherent recombination has primarily focused on limitations stemming from longitudinal motion \cite{margalit_analysis_2019, zimmermann_t3_interferometer_2018}, although consideration has been given to the role of off-axis fields for atomic beam SGIs (see Fig. \ref{fg:sg_apparatus}) \cite{shortdoi:gsmsgh} as well as wave-packet evolution in harmonic and inverted harmonic potentials, both in the context of SGIs \cite{shortdoi:g9rwcz} and a non-recombining Stern-Gerlach experiment \cite{shortdoi:d9pb92}. 
\markstopchange

We begin in Section \ref{se:imagined_experiments} by outlining a proposed SGI using Rb Rydberg atoms.  This proposed SGI will serve as a specific example of the general results.  In Section \ref{se:building_blocks} we indicate the procedure by which we derive expressions for SGI fringe visibilities.  Section \ref{se:different_sgi_vis} contains our primary result: fringe visibilities can be strongly influenced by off-axis fields and with the extent strongly depending on the particular experimental sequence. We close in Section \ref{se:discussion} with a discussion of the relationship between our results and further possible work.

\section{Imagined Electric SGI using cold Rb Rydberg atoms}
\label{se:imagined_experiments}

\begin{figure*}
\begin{center}
\includegraphics{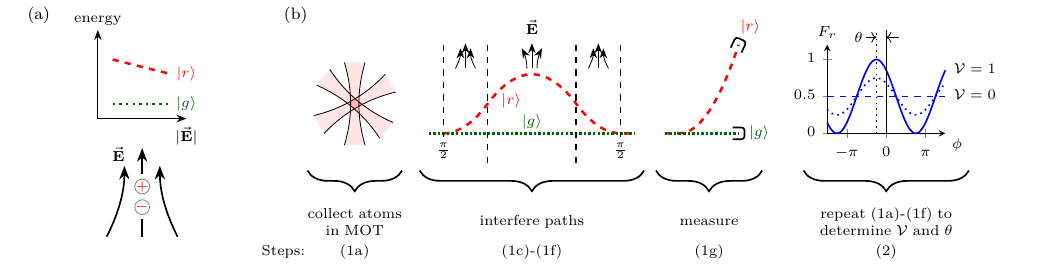}
\end{center}
\caption{\label{fg:experimental_sequence} (a) The shifts of the energies of internal states $\ket{r}$ and $\ket{g}$ as a function of electric field magnitude $|\vect{E}|$ together with an illustration of the origin of the net force on an atom in state $\ket{r}$, and (b) the steps in proposed electric SGI experiment using laser-cooled Rb atoms excited to Rydberg states.  After repeating steps (1a) through (1g), with different phases $\phi$ of the final $\pi/2$ pulse, a fringe visibility $\mathcal{V}$ and phase shift $\theta$ may be determined.
}
\end{figure*}

\subsection{Overview}
\label{se:imagined_overview}

Although there have been reports of matter-wave interferometry using Rydberg atoms \cite{shortdoi:gf432r, shortdoi:gg2bbk, shortdoi:gmsf2r}, two of us have disputed the interpretation of these experiments \cite{shortdoi:gtddkw}, arguing that the observed interferometric phase $\theta$ does not depend on acceleration of the atoms.  When considered together with the severe complications found by Comparat \cite{shortdoi:gptnmb}, it is helpful to know that a Rydberg atom matter-wave interferometry experiment may, in fact, be practical. While such an electric SGI might not offer improvements in measuring $g$, it may be useful for the measurement of electric dipole moments and field gradients.

A complicating factor in Refs \cite{shortdoi:gf432r, shortdoi:gg2bbk, shortdoi:gmsf2r} is that the atoms were in a beam moving at $\SI{2000}{m/s}$ in the lab frame, which would be avoided in an electric SGI consisting of the following steps (see Fig.~\ref{fg:experimental_sequence}(b)):
\begin{enumerate}[(1),nosep]
\item \label{it:to_repeat}
\begin{enumerate}[(a),nosep]
\item \label{it:mot}
collect translationally cold Rb atoms in a vapor cell magneto-optical trap (MOT) \cite{shortdoi:fjh9zv}.
\item \label{it:switch_off}
switch off the gradient magnetic field and laser beams required for the MOT.
\item \label{it:first_pid2}
with a resonant $\pi/2$ coupling pulse, create a superposition between the ground state $\ket{g}$ and a Rydberg state $\ket{r}$.
\item \label{it:sep}
use spatially-varying electric fields to accelerate the $\ket{r}$ state away from the $\ket{g}$ state (the difference in forces being due to their different dipole moments).
\item \label{it:recombine}
recombine the spatial wave-functions for the $\ket{r}$ and $\ket{g}$ states using spatially-varying electric fields.
\item \label{it:second_pid2}
apply a $\pi/2$ coupling pulse identical to that in step \ref{it:first_pid2}, but with a shifted phase $\phi$.
\item \label{it:measure}
measure the relative populations of the Rydberg  $P_{r}(\phi)$
and ground $P_{g}(\phi)$ states to determine the fraction of atoms
that are in either state ($F_{r}$ and $F_{g}$ respectively).
\end{enumerate}
\item \label{it:repeat}
repeat the steps in \ref{it:to_repeat}, with a different relative phase $\phi$ between the coherent couplings used in steps \ref{it:first_pid2} and \ref{it:second_pid2}.
\end{enumerate}
Consider perfect recombination: as the relative phase $\phi$ is varied, we expect $F_{r}$ and $F_{g}$ to oscillate between zero and one, 180 degrees out of phase with each other --- analogous to the two output ports of an optical Mach-Zehnder interferometer (MZI).   Imperfect recombination, the subject of this paper, can be characterized by the standard definition of interferometric visibility \cite{isbn:9781108477437}:
$\mathcal{V} \coloneqq (F_\text{max}-F_\text{min})/(F_\text{max}+F_\text{min})$,
where $F_\text{min}$ and $F_\text{max}$, are the maximum and minimum state fractions of either state --- it doesn't matter which since $F_{g}+F_{r} =1$.  Perfect recombination corresponds to $\mathcal{V}=1$ whereas $\mathcal{V}=0$ signifies completely unsuccessful recombination.

Some details will be given later:
which states should be used (Section \ref{se:edipoles} and Appendix \ref{se:rydberg_details})?
How should the spatially-varying electric fields be generated (Section \ref{se:iefields})?
By what criteria should we consider that Humpty-Dumpty has fallen apart (Section \ref{se:destruction_by_sg})?

Most importantly, we have omitted some details of steps \ref{it:to_repeat}\ref{it:sep} and \ref{it:to_repeat}\ref{it:recombine}. Specifically, there are a variety of ways of varying the field gradients and  inserting one or more $\pi$ state-swaps in the sequence to implement the interferometry (see Section \ref{se:different_sgi_vis}). The differences between the $\mathcal{V}$s for these various sequences are our primary result.

There are also details and variations which do not change our analysis, or the intrinsic feasibility of the experiment, in any substantial way:  for example instead of forming a superposition of the ground and a Rydberg state by coherent optical excitation, one could excite to a Rydberg state and then use microwaves to create coherent superpositions of two Rydberg states: one with a large electric dipole moment and another with a significantly smaller moment. After the last $\pi/2$ beam-combining pulse the relative populations of the two Rydberg states could be measured by selective field ionization (SFI) \cite{isbn:9780521385312}.  This variant would significantly lessen the technical requirements for the optical source used for Rydberg excitation --- since the coherent superpositions of internal states will be formed with microwave synthesizers after optical excitation; e.g., Ref.~\cite{shortdoi:ggz44h}.

\subsection{Electric dipoles}
\label{se:edipoles}

Although electric and magnetic SGIs both depend on the differing accelerations of internal states caused by spatially-varying fields, there is one important difference: while atoms may exhibit magnetic dipole moments in zero magnetic field, they do not normally have permanent electric dipole moments in zero electric field. Instead, we must \emph{polarize} them using an electric field.

The extent and even the orientation of the resulting electric dipole moment (with respect to the polarizing field) depends on the particular atomic state.  In that way, different choices for the two internal SGI states can result in qualitatively different behaviour in a gradient field. For example, depending on the sign of the dipole moment, the two internal states can accelerate in the same direction with different magnitudes or in opposite directions with the same or different magnitudes.

For simplicity, we will focus on the case for which an electric field has been applied that polarizes one of the two internal states (a Rydberg state); the other internal state will have a negligible electric dipole moment in the same field; e.g., the ground electronic state or a much less polarizable Rydberg state. We denote the two states by $\ket{r}$ and $\ket{g}$ respectively.

We refer to the polarizing field in the center of the apparatus as $\vect{E}_0$.
To determine the Stern-Gerlach forces we must know how the internal energy of the atom changes as the position-dependent field magnitude $E =|\vect{E}|$ deviates from $E_0$. Our analysis is based on characterizing the shifts in the energy of the $j$th state ($r$ or $g$) by:
\begin{equation} \label{eq:linear_stark_shift}
W_j(E) \approx W_j\left(E_0\right)- \mu_j \left (E-E_0\right),
\end{equation}
where $\mu_j$ refers to the dipole moment of state $j$.  We consider $\mu_j$ to be a signed quantity, with $\sgn \mu_j$ indicating its orientation with respect to the applied field.  Specifically, $\sgn \mu_j=+1/\!\!-\!\!1$ corresponds to a red/blue state that decreases/increases in energy with electric field magnitude, and is a high/low-field seeker.

For the proposed Rb Rydberg SGI, we chose $\ket{r}$ to be the red-most $n=52$ Rydberg Stark state of Rb in a dc electric field of $E_0 = \SI{210}{V/m}$, where it has an electric dipole moment of $\mu_r \approx 3900 \: q a_0$ (this choice of state and calculation of its dipole moment is justified in Appendix \ref{se:rydberg_details}). By contrast, the ground state electric dipole moment is $\mu_g \approx \SI{e-7}{} q a_0$ for the same $E_0$ (calculated using the polarizabilities from Ref.~\cite{shortdoi:ccsc8n}). Thus if the ground state is chosen for $\ket{g}$ its acceleration can be neglected.

Alternately, as discussed at the end of the previous section, we could choose $\ket{g}$ to be another Rydberg state with a significantly smaller dipole moment than $\ket{r}$.  One possibility is the $54d_{3/2}$ state, for which $\mu_{54d_{3/2}} \approx 370 \: q a_0$  at $E_0 = \SI{210}{V/m}$ (see Appendix \ref{se:rydberg_details}).  Since $\mu_r \approx 10 \mu_{54d_{3/2}}$, we expect that using $54d_{3/2}$ as the $g$ state would cause our main results to only mildly differ quantitatively from $\mu_{g}=0$, and thus we focus on the first case for simplicity.

\markstartchange
As we see later (see Eq.~\ref{eq:eff_potential}), the choice of a red, high-field seeking Rydberg state makes the effective transverse potential that of an \emph{inverted} harmonic oscillator, as opposed to the case of a blue-seeking state, which leads to a normal non-inverted harmonic oscillator potential. For concreteness, we have chosen to focus on red-shifting states throughout, as we think this might be the more common scenario; e.g., Ref.~\cite{shortdoi:dfv4zd}.  This is partially because non-core penetrating Rydberg states that have small positive quantum defects adiabatically connect to the red-most part of the Stark manifold. But, one could also consider an electric SGI with blue-shifted low-field seeking states; see the consequences for transverse visibilities at end of Section \ref{se:transverse_analysis}.
\markstopchange

\subsection{Spatially-varying electric fields}
\label{se:iefields}

Voltages applied to electrodes will create the necessary electric fields and field gradients at the location of the atoms in center of the vacuum chamber.  As with Comparat \cite{shortdoi:gptnmb}, we consider a cylindrically symmetric geometry, so that the electrostatic potential $V$ that gives the field by $\vect{E} = - \vect{\nabla} V$ can be written as (e.g., Ref.~\cite{isbn:9781009397759}):
\begin{equation}\label{eq:leg_exp}
V(r,\theta) = \sum_{\ell=0}^{\infty} A_{\ell} \: r^{\ell} \: P_{\ell} (\cos \theta)
\end{equation}
where $r$ and $\theta$ are the normal radial and polar spherical coordinates with the origin at the center of the apparatus, and the $P_{\ell}$'s are the Legendre polynomials: $P_0(\cos \theta)=1$, $P_1(\cos \theta)= \cos \theta$, $P_2(\cos \theta)=\left[3 \cos^2 \theta -1\right]/2$, etc. Only the $A_1$ and $A_2$ terms are required to give the expansion of the components of $\vect{E}$ up to first order in the spatial coordinates:
\begin{equation}
\vect{E}(x,y,z) = \left(-A_1-2 A_2 z\right) \uvect{z}+A_2 x \: \uvect{x}+A_2 y \: \uvect{y},
\end{equation}
so that $\vect{E}|_0 = E_z|_0 \uvect{z}$, with $E_z|_0 = -A_1$, where $\ldots|_0$ indicates evaluation at the origin.  Atoms at the origin in state $i$ will align with $\vect{E}|_0$ and experience a force due to variation of $E_z$ with $z$; i.e., $\vect{F}_i = \vect{\nabla} (\vect{\mu}_i \cdot \vect{E}) = \mu_i \partial_z E_z |_0 \: \uvect{z}$, where  $\partial_z E_z |_0 = - 2 A_2$.

To understand the off-axis behavior of the SGI we must consider the spatial variations of $E$, as these variations determine the potential that the atoms experience (see Eq.~\ref{eq:linear_stark_shift}).  We ignore the higher-order terms in Eq.~\ref{eq:leg_exp} with $\ell \ge 3$ as they do not contribute to either $E_z|_0$ or $\partial_z E_z |_0$, and they can be made arbitrarily small compared to the $\ell=1$ and $\ell=2$ terms by simply increasing the sizes and separations of the electrodes.  However, even after dropping all but the $A_1$ and $A_2$ terms, the spatial dependence of $E$ is awkward:
\begin{equation} \label{eq:e_field_variation_full}
E =\left|A_1\right|\left[1+4 \frac{A_2}{A_1} z+4 \frac{A_2^2}{A_1^2} z^2+\frac{A_2^2}{A_1^2}\left(x^2+y^2\right)\right]^{1 / 2}.
\end{equation}
But provided that we consider distances away from the origin that are small compared to a characteristic length $d \coloneqq |A_1/A_2| = 2 |E_z|_0 / \partial_z E_z |_0|$, we may approximate Eq.~\ref{eq:e_field_variation_full} as:
\begin{equation} \label{eq:e_field_variation_approx}
E=|A_1|\left[1+2 \epsilon \frac{z}{d}+\frac{1}{2}\frac{\left(x^2+y^2\right)}{d^2}+\cdots\right]
\end{equation}
to second-order in the Cartesian coordinates, where $\epsilon$ is the sign of $A_2/A_1$. The term linear in $z$ is fundamental to the SGI, whereas the $x^2$, $y^2$ and higher-order terms are in essence nuisances. Note the absence of a $z^2$ term.

\begin{figure}
\includegraphics{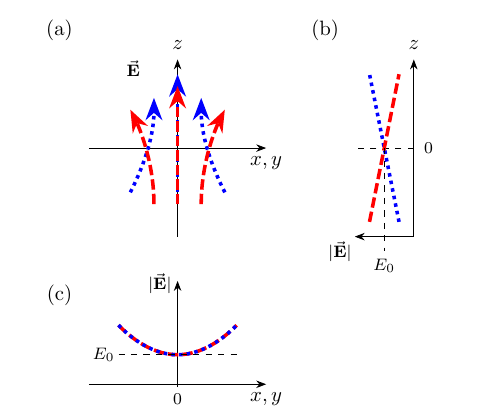}
\caption{\label{fg:field_variations}
(a) Spatial variations of the electric field in a cylindrically symmetric geometry with: $\partial_z E_z |_0 > 0$ (dotted blue line), and $\partial_z E_z |_0 < 0$ (dashed red line); (b) Corresponding variations in the magnitude of the electric field $E$ along the longitudinal $\uvect{z}$ axis and (c) along the transverse $\uvect{x}$ and $\uvect{y}$ axes.  Note that the transverse variation in $E$ is the same for both $\partial_z E_z |_0 > 0$ and $\partial_z E_z |_0 < 0$.  Thus, an atom in a high-field seeking state will experience an force pushing it off the $z$-axis; i.e., the inverted harmonic oscillator potential of Eq.~\ref{eq:eff_potential}.
}
\end{figure}

Substitution of Eq.~\ref{eq:e_field_variation_approx} into Eq.~\ref{eq:linear_stark_shift} gives the effective potential
\begin{equation} \label{eq:eff_potential}
U_j(\vect{r})=W_j(E_0)-\mu_j E_0\left[2 \epsilon \frac{z}{d}+\frac{1}{2}\frac{(x^2+y^2)}{d^2}\right]
\end{equation}
that an internal state $j$ experiences as a function of position.  Note that for red, high-field seeking states ($\sgn \mu_j = 1$) the effective potential in the $\uvect{x}$ and $\uvect{y}$ directions is that of an \emph{inverted} harmonic oscillator, independent of $\epsilon$; i.e., if the field gradient direction is reversed the transverse potential is still that of an inverted harmonic oscillator (see Fig.~\ref{fg:field_variations}).

Equation \ref{eq:eff_potential} may also be applicable to a magnetic SGI with a cylindrical geometry and linear Zeeman shifts, where $\mu_j$ is now a magnetic-dipole moment magnitude, $E_0$ and $E$ are replaced by $B_0$ and $B$, and the other symbols have analogous meanings.
We note that Zhou \emph{et al.}~\cite{shortdoi:g9rwcz} have considered a magnetic SGI which makes use of an inverted harmonic oscillator potential along the primary acceleration axis.


\subsection{Experimental parameters for an electric SGI using Rb Rydberg atoms}
\label{se:representative_experimental_parameters}

We will now describe the parameters for the Rydberg SGI that will be used for numerical examples of the general results.

\begin{figure}
\includegraphics{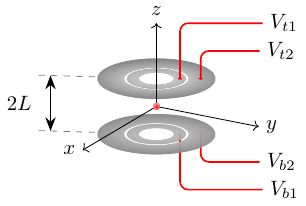}
\caption{\label{fg:electrodes}
An electrode configuration suitable for generating cylindrically symmetric inhomogeneous electric fields.  This arrangement gives reasonable optical access for the MOT and Rydberg excitation beams.  Voltage differences between the inner and outer concentric electrodes ($\Delta V_{\text{grad}} = V_{t1}-V_{t2} = V_{b1}-V_{b2}$) control the field electric field gradient, $\partial_z E_z|_0 \approx \Delta V_{\text{grad}} / L^2$, whereas voltage differences between the top and bottom electrodes set the electric field at the centre of the cloud of atoms.  With the central hole \SI{10}{mm} in diameter and the pairs of electrodes are separated in the $\hat{z}$ direction by $2L = \SI{10}{mm}$, field gradients on the order of $\SI{300e3}{V/m^2}$ are generated by electrode voltage differences $\Delta V_{\text{grad}} \approx \SI{10}{V}$.}
\end{figure}

Comparat \cite{shortdoi:gptnmb} analyzed an SGI sequence involving two internal states with accelerations $a_1$ and $a_2$,
finding that for a total sequence time of $T_{\text{tot}} = 4\tau$, the phase difference between the two arms of the interferometer is expected to be $\theta = m(a_1^2-a_2^2)\tau^3/\hbar$.  We will analyze this same sequence, which we call the \emph{bow sequence}, in more detail later in Section \ref{se:different_sgi_vis}. However, we neglect the acceleration of the $\ket{g}$ state, so that $\theta = m a_r^2 \tau^3/\hbar$.  As a demonstration of matter-wave interferometry, it is desirable to verify this dependence of $\theta$ on $a_r$, and thus we aim for experimental parameters that give a phase shift of at least $\theta \approx 2\pi$.  To avoid significant Rydberg state loss due to radiative effects (see Appendix \ref{se:rydberg_details}), we set $\tau=\SI{1}{\mu s}$, finding then that for $\theta = 2 \pi$ we must have $a_r \approx \SI{68e3}{m/s^2}$.  For $\mu_r \approx 3900 \: q a_0$ (see Section \ref{se:edipoles}), this acceleration requires a field gradient of $\partial_z E_z |_0  \approx \SI{300e3}{V/m^2}$, which is generated with relatively small electrode voltages (see Fig.~\ref{fg:electrodes}).

To determine the influence of $g$, the acceleration due to gravity, we imagine that the longitudinal symmetry axis of the SGI is aligned vertically.  Since the magnitude of $g$ is significantly smaller than our chosen $a_r$, considering a non-zero $g$ results in only a small fractional change in $\theta$.  Specifically, from $\theta = m(a_1^2-a_2^2)\tau^3/\hbar$ with $a_1=a_r+g$ and $a_2=g$ we have $|\Delta \theta / \theta| \approx 2 g/a_r \approx \SI{3e-4}{}$.  Thus, in what follows we do not consider gravity.

We consider the atoms to have a temperature of $T =\SI{100}{\mu K}$, as typically obtained after release from a MOT without further cooling \cite{shortdoi:bgpnv7}.  This temperature corresponds to a thermal de Broglie wavelength of $\lambda =(2\pi\hbar^2/mkT)^{1/2} \approx \SI{20}{nm}$ (here $m$ is the mass of a Rb atom and $k$ is Boltzmann's constant).  We shall see later (in Section \ref{se:different_sgi_vis}) that it is useful that initially stationary atoms will move by $\Delta z = \frac{1}{2} a \tau^2 \approx \SI{34}{nm}$, which is comparable to $\lambda$.

Table \ref{tb:ryd_summary} summarizes the physical parameters for the proposed Rb Rydberg SGI.

\begin{table}
\caption{\label{tb:ryd_summary}
Summary of parameters for a proposed electric SGI, using $^{87}$Rb Rydberg atoms ($n\approx 52$).
}
\begin{ruledtabular}
\begin{tabular}{ccccc}
parameter & & \textrm{SI} &
\markstartchange lin\footnote{Unit system for longitudinal motion (linear potential). See Eq.~\ref{eq:check_def}, which for these parameters gives:
$x_{\egrave}\approx\SI{20}{nm}$
$p_{\egrave}\approx\SI{5.3e-27}{kg \cdot m /s}$, and
$t_{\egrave}\approx\SI{0.54}{\mu s}$.},
$\grave{\phantom{x}}$ &
iho\footnote{Unit system for transverse motion (inverted harmonic potential). See Eq.~\ref{eq:breve_def},
which for these parameters gives:
$x_{\ebreve}\approx\SI{0.39}{\mu m}$
$p_{\ebreve}\approx\SI{2.73e-28}{kg \cdot m /s}$, and
$t_{\ebreve}\approx\SI{0.20}{ms}$.
} , $\breve{\phantom{x}}$ \markstopchange \\
\colrule
$\mu_r$ & 3900 $q a_0$ & & &\\
$E_0$ &  & \SI{210}{V/m} & &\\
$\partial_z E_z|_0$ & & \SI{300e3}{V/m^2} & & \\
$d$ & & \SI{1.4}{mm} & 71000 & 3700 \\
$a_r$ & & \SI{68e3}{m/s^2} & 1 & 7300 \\
$\tau$ & & \SI{1}{\mu s} & 1.8 & \SI{0.0049}{} \\
$\sigma_{\parallel}$, $\sigma_{\perp}$ & & \SI{100}{\mu m} & 5000 & 260 \\
$\lambda$ & for $T=\SI{100}{\mu K}$ & \SI{19}{nm} & 0.94 & 0.048 \\
\end{tabular}
\end{ruledtabular}
\end{table}

\section{Methodology for modeling SGI interferometers}
\label{se:building_blocks}

\subsection{Introduction}

We now consider how to compute the SGI visibilities using an approach similar to that used by Zimmermann \emph{et al.}~\cite{shortdoi:gpb7cs} in one spatial dimension.  Due to the relatively simple form of both the longitudinal ($\uvect{z}$) and transverse ($\uvect{x}$ and $\uvect{y}$) potentials given by Eq.~\ref{eq:eff_potential}, visibilities may be determined according to quantum mechanics without the use of semi-classical methods; e.g., Ref.~\cite{shortdoi:gmsf2r}.

\markstartchange
\subsection{Initial state density matrix}
\label{se:dens_intro}

At the start of the interferometry sequence, we describe the initial \emph{translational} state of the atoms using a density operator that incorporates both their spatial distribution and temperature.  We assume that we are far from quantum degeneracy and neglect any interatomic interactions so that a single-particle density operator suffices.  In this section we provide a rationale for its form.

Since all three spatial dimensions will be treated similarly, we consider one initially ($x$) and assume that the atoms are centered at the origin with an uncertainty of $\sigma_x$.  In the pure-state limit ($T=0$), we take each atom's position space wave-function to be of the form:
\begin{equation} \label{eq:pure_dens_form}
\braket{x|\psi} =
\left(
\frac{1}{\sqrt{2\pi} \sigma_x}
\right)^{1/2}
\exp\left(-\frac{x^2}{4\sigma_x^2} \right).
\end{equation}
For a pure state $\ket{\psi}$, the density operator is $\hat{\rho} = \ket{\psi}\bra{\psi}$, and thus in position space we have:
\begin{align} \nonumber
\braket{x'|\,\hat{\rho}\,|x''} &=  \braket{x'|\psi} \braket{\psi|x''} \\
& =  \label{eq:spatial_pure} \frac{1}{\sqrt{2\pi} \sigma_x}
\exp\left(-\frac{x'^2+x''^2}{4\sigma_x^2} \right).
\end{align}
Note that by explicit integration, we can verify that $\sigma_x = \sqrt{\braket{(x-\braket{x})^2}}$.  Likewise, from Eq.~\ref{eq:spatial_pure} we find that $\sigma_p = \sqrt{\braket{(p-\braket{p})^2}} = \hbar/(2\sigma_x)$, and thus $\sigma_x \sigma_p = \hbar/2$, demonstrating that this pure state saturates the Heisenberg inequality (as is well-known; see e.g., Ref.~\cite{isbn:9781108473224}).

To account for the non-zero temperature of the atoms, we first consider the density matrix arising from quantization in an artificial box of width $L \rightarrow \infty$, with periodic boundary conditions; e.g., Ref.~\cite{isbn:9780201360769}:
\begin{equation} \label{eq:feynman_dens_position}
\braket{x'|\,\hat{\rho}\,|x''} = \frac{1}{L} \exp\left( -\frac{\pi (x'-x'')^2}{\lambda^2}\right).
\end{equation}
The corresponding momentum uncertainty is $\sigma_p  = \sqrt{2\pi} \hbar / \lambda = \sqrt{m k T}$, as expected from the equipartition theorem in one dimension.

Both the spatial distribution of our atoms \emph{and} their non-zero temperature can be accounted for by combining Eq.~\ref{eq:spatial_pure} with \ref{eq:feynman_dens_position}, to obtain our initial state density matrix:
\begin{equation} \label{eq:combined_dens_pos}
\braket{x'|\,\hat{\rho}\,|x''}
= \frac{1}{\sqrt{2\pi}\sigma_x} \exp\left( -\frac{\pi (x'-x'')^2}{\lambda^2}-\frac{x'^2+x''^2}{4\sigma_x^2} \right),
\end{equation}
which is normalized so that $\int_{-\infty}^{\infty} dx \, \rho(x,x) = 1$.
As with the pure-state described by Eq.~\ref{eq:spatial_pure}, the corresponding position uncertainty is still $\sigma_x$, but the momentum uncertainty may be determined from Eq.~\ref{eq:combined_dens_pos} to be:
\begin{equation} \label{eq:sigma_p_gen}
\sigma_p = \frac{\hbar}{2} \sqrt{\frac{8\pi}{\lambda^2} + \frac{1}{\sigma_x^2}},
\end{equation}
which, for $\lambda \rightarrow \infty$, reduces to the pure-state limit $\sigma_p = \hbar/2\sigma_x$, and for $\sigma_x \rightarrow \infty$ reduces to the infinite cloud size limit $\sigma_p = \sqrt{2\pi} \hbar / \lambda$, consistent with Eq.~\ref{eq:feynman_dens_position}.

Rearranging Eq.~\ref{eq:sigma_p_gen} to obtain:
\begin{equation} \label{eq:lambda_given_sigmas}
\lambda = 2 \sigma_x \sqrt{2 \pi / (4 \sigma_x^2 \sigma_p^2 / \hbar^2 -1)}
\end{equation}
allows us to write the initial state density matrix given by Eq.~\ref{eq:combined_dens_pos} in terms of $\sigma_x$ and $\sigma_p$ alone:
\begin{equation} \label{eq:density_matrix_with_sigma_p}
\braket{x'|\,\hat{\rho}\,|x''}
= \frac{1}{\sqrt{2\pi}\sigma_x} \exp\left(
-\frac{(x'+x'')^2}{8\sigma_x^2}
-\frac{(x'-x'')^2}{2\hbar^2/\sigma_p^2}
\right).
\end{equation}
Despite the pleasing symmetry of this form, we express our results using $\lambda$ and $\sigma_x$, taking the three-dimensional generalization of Eq.~\ref{eq:combined_dens_pos} as our initial state. We choose the ($\lambda$, $\sigma_x$) parameterization instead of the ($\sigma_p$, $\sigma_x$) parameterization since both $\lambda$ and $\sigma_x$ can be varied independently in an experiment, whereas $\sigma_p$ depends on $\sigma_x$ (Eq.~\ref{eq:sigma_p_gen}). For example, we can ensure constant $\lambda$ by preparing an internal state of controllable size from a significantly larger cloud of temperature $T$. This spatial selection of $\sigma_x$ is straightforward when both $\ket{r}$ and $\ket{g}$ are Rydberg states, as it can occur during Rydberg excitation using the spatial profiles of the excitation lasers. Alternately, if we choose $\ket{g}$ to be a ground electronic state, this spatial selection could use a stimulated Raman transition between ground electronic-state hyperfine levels using focused laser beams.

Similarly to how Eq.~\ref{eq:combined_dens_pos} and Eq.~\ref{eq:density_matrix_with_sigma_p} can be viewed
as the same density matrix with different parameterizations, in Appendix \ref{se:alt_dens_param} we show that with suitable choices of $\sigma_x$ and $\lambda$, a density matrix of the form of Eq.~\ref{eq:combined_dens_pos} can also describe atoms at a non-zero temperature trapped in a harmonic oscillator potential.  Although we do not make use of this alternative parameterization here, it may be useful for other experiments.

To describe the system just prior to the initial $\pi/2$ beam-splitting pulse at $t=0$ (see Section \ref{se:imagined_overview}), we combine the internal and translational degrees of freedom into the density operator:
\begin{equation}
\hat{\rho}(0^-) = \ket{g}\!\bra{g} \otimes \int d^3\vect{r}\,' d^3\vect{r}\,''
  \rho_0(\vect{r}\,', \vect{r}\,''  ) \ket{\vect{r}\,'}\! \bra{\vect{r}\,''}
\end{equation}
where
\begin{equation} \label{eq:factor_density}
\rho_0(\vect{r}\,', \vect{r}\,'')=\rho_{0,x}(x',x'') \rho_{0,y}(y',y'') \rho_{0,z}(z',z''),
\end{equation}
with $\rho_{0,x} = \braket{x'|\,\hat{\rho}\,|x''}$ as given in Eq.~\ref{eq:combined_dens_pos} and similarly for the $y$ and $z$ dimensions.

\subsection{Calculation of visibilities}

We assume that the duration of the initial $\pi/2$ beamsplitting pulse (and later pulses in the sequence) will be small compared to the other time-scales so that the $\pi/2$ pulse may be considered to act instantaneously. Thus after its application, $\hat{\rho}(0^+) = \hat{U}_{\pi/2} \: \hat{\rho}(0^-) \: \hat{U}^{\dagger}_{\pi/2}$, where
\begin{equation} \label{eq:initial_pid2}
\hat{U}_{\pi/2} = \frac{1}{\sqrt{2}} \left( \ket{g}\bra{g} + \ket{r}\bra{r} - \ket{g}\bra{r} + \ket{r}\bra{g} \right) \otimes \hat{\mathds{I}}.
\end{equation}

\markstopchange

Between pulses coupling the two internal states ($i=g$ or $r$), the system will evolve according to the time-dependent Schr\"{o}dinger equation with the Hamiltonian:
\begin{equation}
\hat{H} = \sum_{j=g,r} \left[
\ket{j}\bra{j}
\otimes
\left(
\frac{\hat{p}^2}{2m} + \int d^3\vect{r} \:\: \Delta U_j(\vect{r}) \: \ket{\vect{r}}\bra{\vect{r}}
\right)
\right]
\end{equation}
where the spatial variations in the potentials $\Delta U_j(\vect{r})$ are given by rearranging Eq.~\ref{eq:eff_potential}; i.e., $\Delta U_j(\vect{r}) = U_j(\vect{r})-W_j(E_0)$.  Note that we are working in a rotating frame, such that the basis vectors for each of the two internal states $j=g,r$ are related to the Schr\"{o}dinger picture kets by $\ket{j} = \ket{j}_S \exp(iW_j(E_0)t/\hbar)$.

To determine the visibility, it is helpful to introduce a matrix representation of the density operators:
\begin{equation} \label{eq:arho}
\hat{\rho}(t) \equiv
\begin{pmatrix}
\rho_{gg}(\vect{r}\,',\vect{r}\,'',t) & \rho_{gr}(\vect{r}\,',\vect{r}\,'',t)\\
\rho_{rg}(\vect{r}\,',\vect{r}\,'',t) & \rho_{rr}(\vect{r}\,',\vect{r}\,'',t)
\end{pmatrix}
\end{equation}
where $\rho_{ij}(\vect{r}\,',\vect{r}\,'',t) = \bra{i} \otimes \bra{\vect{r}\,'} \hat{\rho}(t) \ket{j} \otimes \ket{\vect{r}\,''}$.

In this representation, the final beam-combining $\pi/2$ pulse, phase-shifted by $\phi$ with respect to first $\pi/2$ pulse, takes the form:\footnote{Note that with $\phi=0$, this matrix is the representation of $U_{\pi/2}$ presented in Eq.~\ref{eq:initial_pid2}.}
\begin{equation} \label{eq:matrix_pid2}
\hat{U}_{\pi/2}(\phi) \equiv \frac{1}{\sqrt{2}}
\begin{pmatrix}
1 & -e^{i\phi}\\
e^{-i\phi}  & 1
\end{pmatrix}
\end{equation}
where, as described in Section \ref{se:imagined_overview}, $\phi$ will be varied and resulting oscillations in the final state populations will be used to determine the fringe visibility $\mathcal{V}$.

Immediately after the final $\pi/2$ pulse, at $t_{\pi/2}^+$, the Rydberg population depends on the density matrix just prior to application of the pulse, at $t_{\pi/2}^-$.  Specifically, the probability density for an atom being in the state $\ket{r}$ after application of the final $\pi/2$ pulse is:
\begin{equation} \label{eq:before_trace}
\rho_{rr}(\vect{r},\vect{r},t_{\pi/2}^+) = \frac{1}{2} \left[
1 +
2 \operatorname{Re}\:\{\rho_{rg}(\vect{r}\,',\vect{r}\,'',t_{\pi/2}^-) e^{i\phi}\}
\right].
\end{equation}
If we only observe the internal state populations, we must trace over the unobserved positions of the atoms:
$\rho_{rr}(t_{\pi/2}^+) = \int d^3 \vect{r} \: \rho_{rr}(\vect{r},\vect{r},t_{\pi/2}^+)$.
Distributing this trace over Eq.~\ref{eq:before_trace} and introducing $\theta$ such that $\rho_{rg}(t_{\pi/2}^-) = |\rho_{rg}(t_{\pi/2}^-)| e^{i\theta}$ gives the probability of finding an atom in the $\ket{r}$ state:
\begin{equation}
\rho_{rr}(t_{\pi/2}^+)
=
\frac{1}{2} \left[
1 +
\mathcal{V} \cos (\phi + \theta) \right],
\end{equation}
where the visibility is given by $\mathcal{V}=2 | \rho_{rg}(t_{\pi/2}^-)|$ with
\begin{equation}
\rho_{rg}(t_{\pi/2}^-) = \int d^3 \vect{r} \: \rho_{rg}(\vect{r},\vect{r},t_{\pi/2}^-).
\end{equation}
The computation of $\rho_{rg}(t_{\pi/2}^-)$, and thus $\mathcal{V}$, is simplified for the SGIs of interest here because we satisfy two conditions:
\begin{enumerate}[label=(C\arabic*),nosep]
\item between the initial beam-splitting and final beam-combining $\pi/2$ pulses there are two \emph{distinct} paths through the interferometer.  This is true by design for the sequences that we shall consider.
\item the motion of the atoms in the longitudinal and transverse directions may be considered separately, resulting in the factorization of $\rho_{rg}(t_{\pi/2}^-)$, and thus $\mathcal{V}$, into transverse and longitudinal components, each of which can be computed individually.
\end{enumerate}
Imperfections invalidating (C1), such as imprecise $\pi$ and $\pi/2$ pulses, are largely a matter of experimental implementation and are not expected to be important here.  The validity of (C2) depends on the separability of the Hamiltonian into $x$, $y$, $z$ contributions and critically depends on the validity of Eq.~\ref{eq:eff_potential} for the effective potential $U_j(\vect{r})$.  This in turn depends on the validity of the expansion that describes how the electric field magnitude $E$ varies in space (Eq.~\ref{eq:e_field_variation_approx}) and how the internal energy levels depend on $E$ (Eq.~\ref{eq:linear_stark_shift}) (see Section \ref{se:discussion}).

Condition (C1), the existence of two distinct paths, implies that the net effect of time evolution between the initial $\pi/2$ beam-splitting pulse up to the final $\pi/2$ beam-splitting pulse is diagonal in our matrix representation (Eq.~\ref{eq:arho}). Specifically, for unitary time evolution from just after the initial $\pi/2$ pulse until just before the final $\pi/2$ pulse --- which we write as $\hat{U}(t_{\pi/2}^-,0^+)$ --- the only non-zero matrix elements are $U_{gg}(\vect{r}',\vect{r}'',t_{\pi/2}^-,0^+)$ and $U_{rr}(\vect{r}',\vect{r}'',t_{\pi/2}^-,0^+)$.  Abbreviating these as $\mathcal{U}_g(\vect{r}',\vect{r}'')$ and $\mathcal{U}_r(\vect{r}',\vect{r}'')$ respectively, we have:
\begin{multline} \label{eq:mrhom}
\rho_{rg}(t_{\pi/2}^-) = \frac{1}{2} \int d^3\vect{r}' d^3\vect{r}'' d^3\vect{r}''' \:
\mathcal{U}_r(\vect{r}',\vect{r}'')
\rho_0(\vect{r}'',\vect{r}''')\\
\times \mathcal{U}_g^{\dagger}(\vect{r}''',\vect{r}')
\end{multline}
It is helpful to adopt an operator notation to avoid writing out these explicit integrations and introduce the ``complex visibility'' $V \coloneqq \mathcal{V} e^{i\theta} = 2 \rho_{rg}(t_{\pi/2}^-)$, so that:
\begin{equation} \label{eq:xxxx}
\rho_{rr}(t_{\pi/2}^+) = \frac{1}{2} \left[
1 + \operatorname{Re}\:\{V e^{i\phi} \}
\right]
\end{equation}
where
\begin{equation} \label{eq:how_we_calc_v}
V = \operatorname{Tr} (
\hat{\mathcal{U}}_g^{\dagger} \hat{\mathcal{U}}_r^{\phantom{\dagger}} \hat{\rho}_0
).
\end{equation}
We have made use of the cyclic property of traces to write this expression in a form that emphasizes that if $\hat{\mathcal{U}}_g^{\dagger} \hat{\mathcal{U}}_r^{\phantom{\dagger}} = e^{i\theta} \: \hat{\mathds{I}}$ then $\mathcal{V}=1$ (with a trivial example being the case of two identical paths; i.e., $\hat{\mathcal{U}}_g = \hat{\mathcal{U}}_r$ and $\theta=0$).  Each of $\hat{\mathcal{U}}_g^{\dagger}$ and $\hat{\mathcal{U}}_r^{\phantom{\dagger}}$ consist of compositions of time-evolution operators, specific to each SGI sequence, representing motion during different parts of the sequence (free-evolution, constant force, harmonic oscillator).  These well-known results are contained in Appendix \ref{se:propagators}.

Sequences may contain internal state swaps ($\pi$-\emph{pulses},
$\hat{U}_{\pi} \coloneqq \hat{U}_{\pi/2}(0)^2$; see Eq.~\ref{eq:matrix_pid2}).
Note that, according to our conventions, the $g$ and $r$ labels in $\hat{\mathcal{U}}_g$ and $\hat{\mathcal{U}}_r$ refer to the internal state \emph{just prior} to the final $\pi/2$ pulse; i.e., at $t^-_{\pi/2}$.

Condition (C2), the separability of the motions in the $x$, $y$, and $z$ directions, permits the factorization: $\hat{\mathcal{U}}_j = e^{i\eta_j} \hat{\mathcal{U}}_{j,x} \hat{\mathcal{U}}_{j,y} \hat{\mathcal{U}}_{j,z}$, where the $e^{i\eta_j}$ phase factor is associated with internal state manipulations; i.e., the $\pi$-pulses, if any. (Only the difference in the phases, $\eta_{r,g} \coloneqq \eta_{r}-\eta_{g}$, is measurable, and will be either $\pi$ or zero depending on whether the sequence contains an odd or even number of $\pi$ pulses respectively.)  Our assumed initial density operator $\hat{\rho}_0$ can be factored similarly (see Eq.~\ref{eq:factor_density}), so that the complex visibility may be written as $V=e^{i\eta_{r,g}} \, V_x V_y V_z$, with
\begin{equation} \label{eq:how_we_calc_v_i}
V_x = \operatorname{Tr} (\hat{\mathcal{U}}_{g,x}^{\dagger} \hat{\mathcal{U}}_{r,x}^{\phantom{\dagger}} \hat{\rho}_{0,x})
\end{equation}
and similarly for the $y$ and $z$ components.  We assume that the $x$ and $y$ axes are equivalent ($\sigma_x = \sigma_y \eqqcolon \sigma_{\perp}$ in Eq.~\ref{eq:combined_dens_pos}) so we write $V=e^{i\eta_{r,g}} \, V_{\perp} V_{\parallel}$, where $V_{\perp}=V_x V_y = V_x^2$ and $V_{\parallel}=V_z$.  Analogously for the magnitudes: $\mathcal{V}=\mathcal{V}_{\perp} \mathcal{V}_{\parallel}$, and for the phases: $\theta = \eta_{r,g} + \theta_{\perp} + \theta_{\parallel}$, where $\theta_{\perp} = 2 \theta_{x}$ and $\theta_{\parallel} = \theta_{z}$.

It is a significant simplification to be able to consider the on-axis and off-axis contributions to the visibilities separately.
In what follows, we will make repeated use of Eq.~\ref{eq:how_we_calc_v_i} to determine interferometric visibilities ($\mathcal{V}_{\perp}$ and $\mathcal{V}_{\parallel}$)) and phases ($\theta_{\perp}$ and $\theta_{\parallel}$).

\subsection{A fully-open sequence --- destruction of coherence by Stern-Gerlach splitting}
\label{se:destruction_by_sg}

\begin{figure}
\begin{center}
\includegraphics{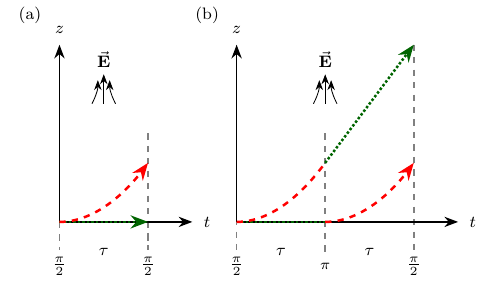}
\end{center}
\caption{\label{fg:two_terrible}
Two interferometry sequences that are open in: (a) both final momentum and position (fully open) and (b) just final position (partially open).  The dashed red line indicates the internal state $\ket{r}$, which accelerates in an inhomogeneous electric field, whereas the dotted green line indicates the internal state $\ket{g}$ which does not.
}
\end{figure}

To illustrate the formalism of the preceding section, imagine the simplest possible SGI (see Fig.~\ref{fg:two_terrible}(a)):  a superposition of the $\ket{r}$ and $\ket{g}$ states is formed with a $\pi/2$ pulse, an inhomogeneous electric field is then applied to accelerate the $\ket{r}$ state for a duration $\tau$, after which a $\pi/2$ beam combining pulse is applied, and finally the $\ket{g}$ and $\ket{r}$ populations are measured.  This sequence would be repeated with different relative phases $\phi$ between the two $\pi/2$ pulses to measure the visibility $V$.

This experiment hardly qualifies as interferometry as classical particles which travel the two paths will not meet in either position or momentum just prior to the final $\pi/2$ pulse.  ``Interferometers'' of this sort are  referred as being \emph{open} \cite{shortdoi:gtrpxw} in either position, momentum, or both (as here).

\emph{Longitudinal visibility and phase:} we determine $V_{\parallel}=\mathcal{V}_{\parallel} e^{i \theta_{\parallel}}$ using Eq.~\ref{eq:how_we_calc_v_i} (replacing $x$ by $z$) with $\hat{\mathcal{U}}_{r,z} = \hat{\mathcal{K}}_{\text{lin}}(\tau,+)$ and
$\hat{\mathcal{U}}_{g,z} = \hat{\mathcal{K}}_{\text{free}}(\tau)$, the time-evolution operators for a particle in a linear potential and a free particle respectively, as given in Appendix \ref{se:propagators}. \markstartchange We find that:
\begin{equation} \label{eq:worst_lvis_SI}
\mathcal{V}_{\parallel} = \exp \left[
-\frac{1}{2}\frac{F^2}{\hbar^2} \sigma_{\parallel}^2 \tau^2
- \frac{F^2}{m^2}  \left( \frac{\pi}{4\lambda^2} + \frac{1}{32\sigma_{\parallel}^2} \right) \tau^4
\right]
\end{equation}
and $\theta_{\parallel} = \tau^3F^2/(3\hbar m)$, where $F$ is the force component of the linear potential defined in Appendix \ref{se:propagators}.
The constants in Eq. \ref{eq:worst_lvis_SI} can be removed by introducing a scaled unit system for the parameters:
\begin{equation} \label{eq:scaled_def}
\grave{\sigma_{\parallel}} \coloneqq \frac{\sigma_{\parallel}}{x_{\egrave}},
\quad
\grave{\lambda} \coloneqq \frac{\lambda}{x_{\egrave}},
\quad
\grave{\tau} \coloneqq \frac{\tau}{t_{\egrave}},
\end{equation}
where $x_{\egrave}$ and $t_{\egrave}$ are defined in Eq. \ref{eq:check_def} within Appendix \ref{se:propagators}. \markstopchange Thus, the visibility can be rewritten as:
\begin{equation} \label{eq:worst_lvis}
\mathcal{V}_{\parallel} = \exp \left[
-\frac{1}{2} \grave{\sigma}_{\parallel}^2 \grave{\tau}^2
- \left( \frac{\pi}{4\grave{\lambda}^2} + \frac{1}{32\grave{\sigma}_{\parallel}^2} \right) \grave{\tau}^4
\right]
\end{equation}
and $\theta_{\parallel} = \grave{\tau}^3/3$.  Using the proposed Rydberg SGI parameters presented in Section \ref{se:representative_experimental_parameters} (see Table \ref{tb:ryd_summary}), we find a drop to $\mathcal{V}_{\parallel}=0.5$  with $\tau = \SI{9}{ns}$, which is overwhelmingly due to the $\exp\left(-\frac{1}{2} \grave{\sigma}_{\parallel}^2 \grave{\tau}^2 \right)$ factor in Eq.~\ref{eq:worst_lvis}.  We will delay a detailed discussion of the transverse behavior until Section \ref{se:different_sgi_vis} (see also Appendix \ref{se:vis_bell}), but for now note that after $\SI{9}{ns}$, the associated transverse visibility drops to $\mathcal{V}_{\perp} \approx 0.3$, giving an indication that transverse motion is significant.

We may gain insight into Eq.~\ref{eq:worst_lvis} by recognizing it as a special case of the ``Humpty-Dumpty'' expression for the visibility of a beam SGI originally obtained by Schwinger \emph{et al.}~\cite{shortdoi:cdxkgn,shortdoi:c34q7x}:\footnote{Equation \ref{eq:hd} is written in the same form found in Margalit \emph{et al.} \cite{shortdoi:gkcvdb}, which differs slightly from the conventions of Ref.~\cite{shortdoi:cdxkgn}.}
\markstartchange
\begin{equation} \label{eq:hd}
\mathcal{V}_{\parallel}
 = \exp \left[
-\frac{1}{2}\left(\frac{\Delta p_z}{\hbar/ \sigma_{z}}\right)^2
-\frac{1}{2} \left(\frac{\Delta z}{\hbar / \sigma_{p,z}}\right)^2
\right],
\end{equation}
where $\Delta z$ and $\Delta p_z$ are the final position and momentum mismatches of the two interferometer arms according to classical mechanics, and $\sigma_z$ and $\sigma_{p,z}$ are the position and momentum uncertainties of the initial state; recall that $\sigma_{p,z}$ is given in terms of $\lambda$ by Eq.~\ref{eq:sigma_p_gen} (we have updated the notation from $\sigma_p$ to $\sigma_{p,z}$). As shown in Ref.~\cite{shortdoi:gpcbfg}, Eq.~\ref{eq:hd} is quite general, applying to any sequence of linear potentials.
\markstopchange

As applied to the fully-open sequence of Fig.~\ref{fg:two_terrible}(a), the $\Delta p_z$ momentum mismatch term in the exponent of Eq.~\ref{eq:hd} corresponds to the $\grave{\tau}^2$ term in the exponent of Eq.~\ref{eq:worst_lvis}, since under constant acceleration the momentum mismatch grows linearly with $\grave{\tau}$. The $\Delta z$ position mismatch term in the exponent of Eq.~\ref{eq:hd} corresponds to the $\grave{\tau}^4$ term in the exponent of Eq.~\ref{eq:worst_lvis} as $\Delta z$ grows like $\grave{\tau}^2$.

We now consider a sequence which eliminates this momentum mismatch and thus also the rapid $\grave{\tau}^2$ contribution to the fall-off in $\mathcal{V}_{\parallel}$.




\subsection{A partially-open sequence --- elimination of longitudinal momentum mismatch}
\label{se:measurement_of_lambda_t}

Final momentum mismatch can be eliminated by accelerating along both paths equally; see Fig.~\ref{fg:two_terrible}(b).  Specifically, after an initial acceleration duration of $\tau$, a $\pi$ pulse can be inserted to swap internal states $\ket{r} \leftrightarrow \ket{g}$, followed by a wait by $\tau$, and then application of a final $\pi/2$ pulse prior to measurement.  The same inhomogeneous electric field is applied throughout both periods $\tau$. This sequence was studied in the Rydberg atom interferometry experiments presented in Ref.~\cite{shortdoi:gmsf2r}.

\emph{Longitudinal visibility and phase:} we determine $V_{\parallel}=\mathcal{V}_{\parallel} e^{i \theta_{\parallel}}$ using Eq.~\ref{eq:how_we_calc_v_i} (replacing $x$ with $z$) with
$\hat{\mathcal{U}}_{r,z} = \hat{\mathcal{K}}_{\text{lin}}(\tau,+) \hat{\mathcal{K}}_{\text{free}}(\tau)$ and
$\hat{\mathcal{U}}_{g,z} = \hat{\mathcal{K}}_{\text{free}}(\tau) \hat{\mathcal{K}}_{\text{lin}}(\tau,+)$
the time-evolution operators for a particle in a linear potential and a free particle, as given in Appendix \ref{se:propagators}.  We find that:
\begin{equation} \label{eq:second_worst_lvis}
\mathcal{V}_{\parallel} = \exp \left[
- \left( \frac{\pi}{\grave{\lambda}^2} + \frac{1}{8\grave{\sigma}_{\parallel}^2} \right) \grave{\tau}^4
\right]
\end{equation}
and $\theta_{\parallel} = \grave{\tau}^3/2$, where $\grave{\sigma}_{\parallel}$, $\grave{\lambda}$, and $\grave{\tau}$ are expressed in the scaled unit system discussed in Appendix \ref{se:propagators}.

Using the parameters given in Table \ref{tb:ryd_summary} (apart from $\tau$), we find that for the proposed electric SGI, the longitudinal visibility drops to $\mathcal{V}_{\parallel} = 0.5$ for $\tau \approx \SI{360}{ns}$, a significant improvement over the ``fully-open'' sequence.  Furthermore, there is a negligible drop in transverse visibility after $\SI{360}{ns}$; i.e., $\mathcal{V}_{\perp} \approx 1.0$.

Examining Eq.~\ref{eq:second_worst_lvis} and noting that for the proposed electric SGI, $\pi/\grave{\lambda}^2 \approx 3.5$, whereas $1/(8\grave{\sigma}_{\parallel}^2) \approx \SI{4.9e-9}{}$, it is apparent that the fall-off in $\mathcal{V}_{\parallel}$ is chiefly due to the $\exp (-\pi \grave{\tau}^4 /\grave{\lambda}^2 )$ factor in Eq.~\ref{eq:second_worst_lvis}.  This dependence suggests that, for this sequence,
a measurement of the decay in the visibility with increasing $\tau$ determines $\lambda$, and thus the temperature, of the sample.  This is consistent with comments \cite{shortdoi:gtddkw} on the results of Ref.~\cite{shortdoi:gmsf2r}, which note that although variations in the phase of the observed fringes did not depend on the acceleration of the Rydberg atoms, the observed reduction in phase contrast could be attributed to the mismatches in final position comparable to $\lambda$.

Both the ``fully-open'' and ``partially-open'' sequences offer insight into the transverse visibility results for more complicated sequences, as we shall now discuss.

\section{Different SGI sequences and their visibilities}
\label{se:different_sgi_vis}


\subsection{Longitudinal analysis}

\begin{figure*}
\includegraphics{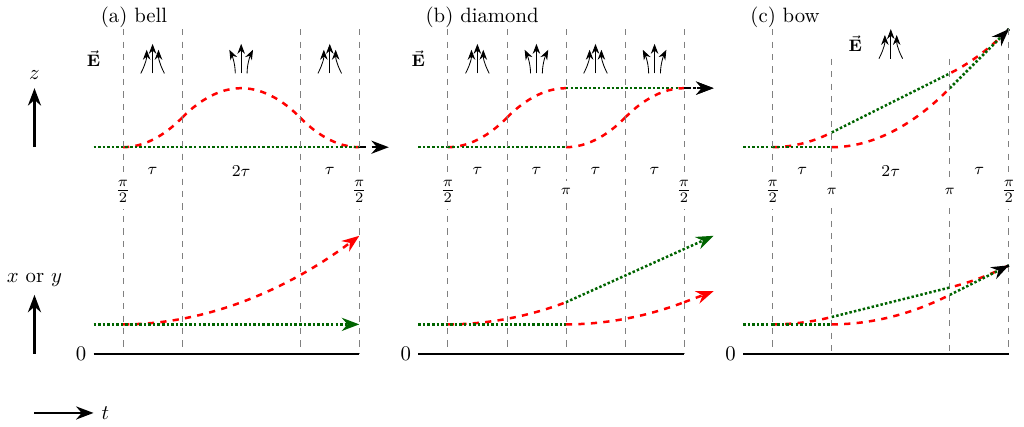}
\caption{\label{fg:longitudinal_trajectories}
Classical trajectories for three different SGI sequences: (a) bell, (b) diamond, and (c) bow.  The dashed red line indicates the internal state $\ket{r}$, which accelerates in an inhomogeneous electric field, whereas the dotted green line indicates the internal state $\ket{g}$, which does not.  The top panels indicate motion in the longitudinal $z$ direction, whereas the bottom panels show the transverse trajectory of an atom which is initially displaced off the $z$ (symmetry) -axis. When in the $\ket{r}$ state, the atom is pushed even further off-axis by the inverted harmonic oscillator potential. Although we have shown off-axis motion assuming linear potentials, this is an idealization: in practice, the off-axis bow sequence will not be perfectly closed. See main text.
}
\end{figure*}

There are a variety of SGI sequences that may be considered given the ability to swap internal states (with $\pi$ pulses) and reverse the field gradient direction (and thus acceleration).  We have determined the visibilities for the three different SGI sequences illustrated in Fig.~\ref{fg:longitudinal_trajectories}, which we have colloquially named the \emph{bell}, \emph{diamond}, and \emph{bow} sequences based on the appearance of their associated classical trajectories. These sequences progress in complexity.

\emph{Bell.}  This sequence is closest in spirit to the original Humpty-Dumpty experiment in the sense that it only uses two field gradient reversals and no internal state swaps ($\pi$ pulses).

\begin{figure}
\begin{center}
\includegraphics{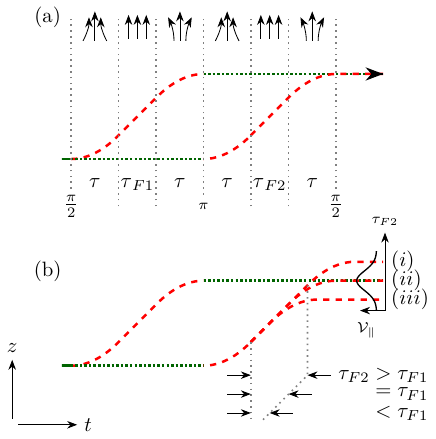}
\end{center}
\caption{\label{fg:open_diamond} (a) Insertion of two free-flight periods $\tau_{F1}$ and $\tau_{F2}$ into the diamond SGI sequence. (b) Variation of the difference between $\tau_{F2}$ and $\tau_{F1}$ to create final position mismatchs and corresponding decreases in longitudinal visibility $\mathcal{V}_{\parallel}$; see Eq.~\ref{eq:diamond_measure_vis} in the main text.
 }
\end{figure}

\emph{Diamond.}  This sequence involves three field gradient reversals and a single $\pi$ pulse.  The combined shape of the two classical paths are reminiscent of a standard photon recoil matter-wave interferometer \cite{shortdoi:ggxx4v}, especially if one inserts free flight segments (see Fig.~\ref{fg:open_diamond}).

\emph{Bow.} This sequence uses two $\pi$ pulses and does not involve any field gradient reversals.  The longitudinal motion along the gradient field direction has been analyzed by several authors \cite{shortdoi:gpb7cs,shortdoi:gt4jhh} with Comparat \cite{shortdoi:gptnmb} also studying the transverse motion.

The interferometric phases and visibilities for all three sequences were determined using the methods outlined in Section \ref{se:building_blocks}.  For longitudinal motion, Table \ref{tb:main_long_results} tabulates the unitary time evolution operators
$\hat{\mathcal{U}}_{r,z}$
and
$\hat{\mathcal{U}}_{g,z}$
associated with each path.  These operators are inserted into Eq.~\ref{eq:how_we_calc_v_i} (replacing $x$ with $z$) to determine the longitudinal visibility
$\mathcal{V}_{\parallel}$ and interferometric phase $\theta_{\parallel}$.  The integrations corresponding to the composition of the density operator with
$\hat{\mathcal{U}}_{r,z}$
and
$\hat{\mathcal{U}}_{g,z}$
together with the trace in Eq.~\ref{eq:how_we_calc_v_i} are performed using computer algebra \cite{github:our_repo}.

\begin{table*}
\caption{\label{tb:main_long_results}
Longitudinal motion time-evolution operators for each path and resulting phases for different SGI sequences in a cylindrical geometry.
}
\begin{ruledtabular}
\begin{tabular}{cccc}
sequence & $\hat{\mathcal{U}}_{r,z}$ & $\hat{\mathcal{U}}_{g,z}$ & $\theta_{\parallel} = \theta_z$ \\[1mm] \hline
& & & \\[-3mm]
bell
&
\begin{math}
\hat{\mathcal{K}}_{\text{lin}}(\grave{\tau},+)
\hat{\mathcal{K}}_{\text{lin}}(2\grave{\tau},-)
\hat{\mathcal{K}}_{\text{lin}}(\grave{\tau},+)
\end{math}
&
\begin{math}
\hat{\mathcal{K}}_{\text{free}}(4\grave{\tau})
\end{math}
&
$-\frac{2}{3} \grave{\tau}^3$
\\
diamond\footnote{%
To measure temperature by opening the diamond sequence up in position, we can insert two free-flight segments of duration $\tau_{F1}$ and $\tau_{F2}$ (see Fig.~\ref{fg:open_diamond}), so that:\\
\begin{math}
\hat{\mathcal{U}}_r =
\hat{\mathcal{K}}_{\text{lin}}(\grave{\tau},-)
\hat{\mathcal{K}}_{\text{free}}(\grave{\tau}_{F2})
\hat{\mathcal{K}}_{\text{lin}}(\grave{\tau},+)
\hat{\mathcal{K}}_{\text{free}}(\grave{\tau}+\grave{\tau}_{F1}+\grave{\tau})
\end{math}
and
\begin{math}
\hat{\mathcal{U}}_g =
\hat{\mathcal{K}}_{\text{free}}(\grave{\tau}+\grave{\tau}_{F2}+\grave{\tau})
\hat{\mathcal{K}}_{\text{lin}}(\grave{\tau},-)
\hat{\mathcal{K}}_{\text{free}}(\grave{\tau}_{F1})
\hat{\mathcal{K}}_{\text{lin}}(\grave{\tau},+)
\end{math}.
}
&
\begin{math} 
\hat{\mathcal{K}}_{\text{lin}}(\grave{\tau},-)
\hat{\mathcal{K}}_{\text{lin}}(\grave{\tau},+)
\hat{\mathcal{K}}_{\text{free}}(2\grave{\tau})
\end{math}
&
\begin{math} 
\hat{\mathcal{K}}_{\text{free}}(2\grave{\tau})
\hat{\mathcal{K}}_{\text{lin}}(\grave{\tau},-)
\hat{\mathcal{K}}_{\text{lin}}(\grave{\tau},+)
\end{math}
&
0
\\
bow
&
\begin{math}
\hat{\mathcal{K}}_{\text{lin}}(\grave{\tau},+)
\hat{\mathcal{K}}_{\text{free}}(2\grave{\tau})
\hat{\mathcal{K}}_{\text{lin}}(\grave{\tau},+)
\end{math}
&
\begin{math}
\hat{\mathcal{K}}_{\text{free}}(\grave{\tau})
\hat{\mathcal{K}}_{\text{lin}}(2\grave{\tau},+)
\hat{\mathcal{K}}_{\text{free}}(\grave{\tau})
\end{math}
&
$\grave{\tau}^3$
\\
\end{tabular}
\end{ruledtabular}

\caption{\label{tb:main_trans_results}
Transverse motion time-evolution operators for each path and resulting visibilities for different SGI sequences in a cylindrical geometry. \markstartchange The parameters $\breve{\tau}$, $\breve{\sigma}_{\perp}$, and $\breve{\lambda}$ are defined by scaling the corresponding parameters in SI units by the appropriate factor from Eq. \ref{eq:breve_def}. \markstopchange
}
\begin{ruledtabular}
\begin{tabular}{cccc}
sequence &
$\hat{\mathcal{U}}_{r,x}$ & $\hat{\mathcal{U}}_{g,x}$ & $ \mathcal{V}_{\perp} = |V_x|^2$  \\[1mm] \hline
& & & \\[-3mm]
bell
&
\begin{math}
\hat{\mathcal{K}}_{\text{iho}}(4\breve{\tau})
\end{math}
&
\begin{math}
\hat{\mathcal{K}}_{\text{free}}(4\breve{\tau})
\end{math}
&
$1 - 8\,\breve{\sigma}_{\perp}^4 \breve{\tau}^2 + O(\breve{\tau}^4)$
\\[3mm]
diamond
&
\begin{math} 
\hat{\mathcal{K}}_{\text{iho}}(2\breve{\tau})
\hat{\mathcal{K}}_{\text{free}}(2\breve{\tau})
\end{math}
&
\begin{math} 
\hat{\mathcal{K}}_{\text{free}}(2\breve{\tau})
\hat{\mathcal{K}}_{\text{iho}}(2\breve{\tau})
\end{math}
&
$1 - \left(8 + \dfrac{32\pi\breve{\sigma}_{\perp}^2}{\breve{\lambda}^2}\right) \breve{\tau}^4 + O(\breve{\tau}^6)$
\\[3mm]
bow
&
\begin{math}
\hat{\mathcal{K}}_{\text{iho}}(\breve{\tau})
\hat{\mathcal{K}}_{\text{free}}(2\breve{\tau})
\hat{\mathcal{K}}_{\text{iho}}(\breve{\tau})
\end{math}
&
\begin{math}
\hat{\mathcal{K}}_{\text{free}}(\breve{\tau})
\hat{\mathcal{K}}_{\text{iho}}(2\breve{\tau})
\hat{\mathcal{K}}_{\text{free}}(\breve{\tau})
\end{math}
&
\begin{math}
1 - \left(
-2
+ \dfrac{32\pi^2}{\breve{\lambda}^4}
+ \dfrac{1}{2\breve{\sigma}_{\perp}^4}
+ \dfrac{8\pi}{\breve{\lambda}^2\breve{\sigma}_{\perp}^2}
+ 2 \breve{\sigma}_{\perp}^4
\right) \breve{\tau}^6
+  O(\breve{\tau}^8)
\end{math}
\\
\end{tabular}
\end{ruledtabular}

\end{table*}

For all three sequences there is no drop in the longitudinal visibility $\mathcal{V}_{\parallel}=1$ as $\tau$ increases from zero, independent of the values of $\sigma_{\parallel}$ and $\lambda$.  This is expected since all three sequences are closed in the longitudinal direction:  the final positions and momenta are the same for the classical trajectories of both paths (which is not true for the transverse direction, as we shall soon discuss).  There is, however, a non-zero phase $\theta_{\parallel}$ for both the bell and bow sequences, scaling like $\tau^3$ as shown in Table \ref{tb:main_long_results}; i.e., the so-called ``Kennard phase'' \cite{shortdoi:gpb7cs}.

Recall that the experimental parameters given in Table \ref{tb:ryd_summary} have been chosen so that the accumulated phase is $2 \pi$ for the bow sequence; i.e., $\theta_{\parallel} = \grave{\tau}^3 = 2\pi$ in the ``lin'' units discussed in Appendix \ref{se:propagators}; equivalently, in SI units $\theta_{\parallel} = (\tau/ t_{\egrave}\!)^3 = m a_r^2 \tau^3 / \hbar = 2\pi$.  Measuring the increase in this phase as the field gradient, and thus corresponding acceleration $a_r$, is increased from zero would be a strong indication of matter-wave interferometry.

On the other hand, as might be expected from the symmetry between the two paths, the diamond sequence shows zero longitudinal phase; i.e., $\theta_{\parallel}=0$.

Due to its symmetry, a variant of the diamond sequence is useful for illustrating the effect of slightly ``opening'' an interferometric sequence in position.  As illustrated in Fig.~\ref{fg:open_diamond}, two free-flight durations, $\tau_{F1}$ and $\tau_{F2}$, can be inserted into the diamond sequence and their difference used to control the final position mismatch $|\Delta z| = |a_r \: \tau (\tau_{F2}-\tau_{F1})|$ while still matching final momenta. By substitution of the appropriate
$\hat{\mathcal{U}}_{r,z}$
and
$\hat{\mathcal{U}}_{g,z}$ (see Table \ref{tb:main_long_results})
into Eq.~\ref{eq:how_we_calc_v_i}, we find that
\begin{equation} \label{eq:diamond_measure_vis}
V_{\parallel} = \exp \left[
-\frac{1}{8} \left(
\frac{8\pi}{\lambda^2} +
\frac{1}{\sigma_{\parallel}^2}
\right) (\Delta z)^2
\right].
\end{equation}
Analogously to the fully and partially open sequences discussed in Section \ref{se:building_blocks}, this expression for the longitudinal visibility is a variant of the Humpty-Dumpty equation (Eq.~\ref{eq:hd}). Specifically, Eq.~\ref{eq:diamond_measure_vis} indicates that for our proposed electric SGI using Rb Rydberg atoms with $\sigma_{\parallel} \gg \lambda$, measuring the visibility drop while ``opening up'' the sequence could be used to determine $\lambda$ and thereby the temperature of the atoms (see the estimate of $\Delta z$ given in Section \ref{se:representative_experimental_parameters}).  Two free-flight periods could also be inserted into the bow sequence to measure $\lambda$.

Observing sequences with high visibility, where at some point in the sequence $\Delta z > \lambda$ would, in some sense, constitute putting Humpty-Dumpty back together again.

\subsection{Transverse analysis}
\label{se:transverse_analysis}

So far, we have been focusing on the longitudinal contribution $\mathcal{V}_{\parallel}$ to the overall visibility $\mathcal{V}=\mathcal{V}_{\parallel} \mathcal{V}_{\perp}$. Now, we consider the transverse visibility $\mathcal{V}_{\perp}$. For our cylindrical symmetric geometry, $x$ and $y$ are equivalent, so that $\mathcal{V}_{\perp}=|V_x|^2$.

The procedure for computing $V_x$ is similar to that for the longitudinal visibility $V_z$:  we simplify Eq.~\ref{eq:how_we_calc_v_i} using the time evolution operators for the two paths,
$\hat{\mathcal{U}}_{r,x}$
and
$\hat{\mathcal{U}}_{g,x}$,
as tabulated in Table \ref{tb:main_trans_results}.  Note that when in a linear potential in the $z$ direction, the transverse potential is always that of an inverted harmonic oscillator (for high-field seeking states), pushing trajectories away from the $z$ axis ($x=y=0$).

All of the integrals involved in simplifying Eq.~\ref{eq:how_we_calc_v_i} are of the Gaussian form and can thus be calculated analytically; e.g., Eqs.~1.4 and 1.5 of Ref.~\cite{isbn:9780198566755}. However, the results are unwieldy: the expression for the transverse visibility $\mathcal{V}_{\perp}$ of the bell sequence is given in Appendix \ref{se:vis_bell} and code for deriving the expressions is available at \cite{github:our_repo}.

Since for all three sequences, $\mathcal{V}_{\perp}=1$ as $\breve{\tau} \rightarrow 0$, we compare them by expanding in a power series in $\breve{\tau}$ about $\breve{\tau}=0$.  The resulting expansions are shown in the rightmost column of Table \ref{tb:main_trans_results}.  Note that the leading order correction to $\mathcal{V}_{\perp}=1$ scales like $\breve{\tau}^2$ for the bell sequence, $\breve{\tau}^4$ for the diamond sequence, and $\breve{\tau}^6$ for the bow sequence. These dissimilarities indicate that for sufficiently small values of $\breve{\tau}$, the bow sequence will have a higher visibility than the diamond or bell sequences.

\begin{figure}
\begin{center}
\includegraphics{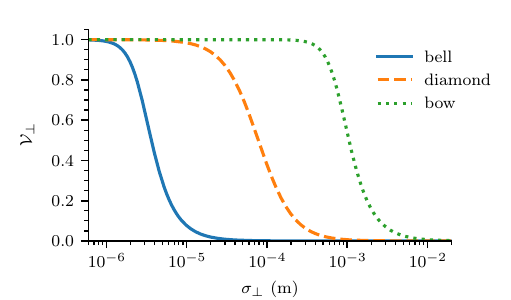}
\caption{\label{fg:vis_versus_sigma_ryd}
Transverse visibility as a function of cloud size for a proposed electric SGI using Rb Rydberg atoms with the parameters given in Table \ref{tb:ryd_summary}.
All calculations use the full expression for $\mathcal{V}_{\perp}$ \cite{github:our_repo}.
}
\end{center}
\end{figure}

To assess if these differences are significant for the proposed SGI, we have evaluated the transverse visibilities as a function of cloud size $\sigma_{\perp}$ for all three sequences, with all other physical parameters as given in Table \ref{tb:ryd_summary}.  As shown in Fig.~\ref{fg:vis_versus_sigma_ryd}, the three sequences are significantly different.  Using the full expression, we can determine the cloud sizes for which the transverse visibility drops to\footnote{The choice of $\mathcal{V}_{\perp}=0.5$ is somewhat arbitrary --- while there is only an improvement of a factor of two in interferometric sensitivity for $\mathcal{V}_{\perp} \approx 1$ \cite{shortdoi:gptnmb}, interferometers which trade much lower visibilities for larger numbers of atoms would require more specific analysis.}  $\mathcal{V}_{\perp}=0.5$. For the bell sequence: $\sigma_{\perp} \approx \SI{4}{\mu m}$; for the diamond sequence: $\sigma_{\perp} \approx \SI{80}{\mu m}$; and for the bow sequence: $\sigma_{\perp} \approx \SI{1}{mm}$.  Achieving these cloud sizes $\sigma_{\perp}$ by focusing optical beams would be straightforward for the bow sequence, but difficult for the bell sequence.  Furthermore, if we wish to avoid interatomic Rydberg-Rydberg interactions by limiting Rydberg density, a smaller cloud size reduces the number of atoms contributing to the measured signal.

For all three sequences, transverse visibility can be increased by lowering the ``frequency'' $\omega$ associated with the inverted harmonic oscillator potential; see Eq.~\ref{eq:omega}.  If $\partial_z E_z|_0$ and $\mu_r$ remain constant, increasing $E_0$ will reduce $\omega$; unfortunately, the maximum value of $E_0$ is limited by the Rydberg Stark structure (see Fig.~\ref{fg:stark_map}).

\markstartchange
Although we focus our analysis on high-field seeking red states (see Section \ref{se:edipoles}), which experience an inverted harmonic transverse potential, our approach may be applied to blue states as well. Such states experience a harmonic transverse potential $\hat{\mathcal{K}}_{\text{ho}}$ , related to the inverted harmonic potential $\hat{\mathcal{K}}_{\text{iho}}$ by the substitution $\omega \rightarrow \omega i$. Thus, the visibility for a blue state may be found using the operator sequences depicted in Table \ref{tb:main_trans_results} with the corresponding substitution of $\hat{\mathcal{K}}_{\text{iho}} \rightarrow \hat{\mathcal{K}}_{\text{ho}}$.

We find that, within the proposed Rydberg SGI, the visibilities for blue and red states are negligibly different --- the blue state version of Fig.~\ref{fg:vis_versus_sigma_ryd} is visually indistinguishable from the red state version displayed here. In fact, for the bell and diamond sequences, the leading order series expansions of $\mathcal{V}_{\perp}$ are exactly the same for both (see Table \ref{tb:main_trans_results}). The series for the bow sequence differs only in the sign of the $-2$ in the $\tau^6$ coefficient, which is positive for a blue state.

\subsection{Physical intuition for differences between sequences}

To obtain an intuitive understanding of the variation in the transverse visibilities for the three sequences, recall the discussion of fully and partially open sequences given in Sections \ref{se:destruction_by_sg} and \ref{se:measurement_of_lambda_t}.  Although these discussions were for motion in the longitudinal direction, let us consider how similar ideas apply to the \emph{transverse} motion of atoms that are initially slightly off-axis and move by small enough distances that the effective potential that they experience can be considered linear.

The bottom panels of Fig.~\ref{fg:longitudinal_trajectories} illustrate the transverse motion of each sequence. Note the similarities to the longitudinal fully and partially open sequences depicted in Fig.~\ref{fg:two_terrible}. In the transverse direction, the bell sequence is fully open: there is a mismatch of both the final position and momentum. In our linear approximation, it is therefore analogous to the longitudinally fully-open sequence considered in Section \ref{se:destruction_by_sg}. The diamond sequence is an improvement, as there is now only a mismatch in final position; i.e., it is partially open. Thus, it may be compared to the sequence considered in Section \ref{se:measurement_of_lambda_t}. Finally, if the atoms experienced an exactly linear potential in the off-axis direction, then the transverse trajectories of the bow sequence would be fully  closed, and there would be no drop in the transverse visibility (recall that we always have $\mathcal{V}_{\parallel}=1$ for longitudinal motion if the sequence is fully closed). However, since the off-axis potential is not linear, the transverse visibility does drop for the bow sequence.

We will now quantitatively compare the transverse visibility of the bell and diamond sequences and the longitudinal visibility of fully and partially open sequences. Specifically, we will use the Humpty-Dumpty equation to approximate the transverse visibilities. In the relevant regime, where we consider the effective potential to be linear, we approximate the potential as $V_{\perp}=-(x+y)F_{\perp}$, with $F=m\omega^2\sigma_{\perp}$. The scale factors of this potential may thus be written:
\begin{equation}
\label{eq:linear_approx_scale_factor_def}
    x_{\egrave}=\left( \frac{x_{\ebreve}^4}{\sigma} \right)^{1/3},
    \quad
    t_{\egrave}=\frac{1}{\breve{\sigma}^{2/3} \omega}
\end{equation}
as defined in Eq. \ref{eq:check_def} and Eq. \ref{eq:breve_def}.

The transverse visibility of the bell sequence for each axis, $\mathcal{V}_{x}$, may be approximated by the longitudinal visibility of the fully-open sequence $\mathcal{V}_{\parallel}$, given by Eq. \ref{eq:worst_lvis}. It is instructive to consider the series expansion of $\mathcal{V}_{\perp}$ in this approximation. To leading order:
\begin{equation}
    \mathcal{V}_{\perp}=|\mathcal{V}_{x}|^2 \approx 1 - 16 \grave{\sigma}_{\perp}^2 \grave{\tau}^2 + O(\grave{\tau}^4)
\end{equation}
reflective of the momentum mismatch factor in Eq. \ref{eq:worst_lvis} with the overall length of the sequence scaled to $4\tau$ for consistency with Table \ref{tb:main_trans_results}. Through substitution of the scale factors with those given in Eq. \ref{eq:linear_approx_scale_factor_def}, this may be rewritten as:
\begin{equation}
    \mathcal{V}_{\perp} \approx 1 - 16 \breve{\sigma}_{\perp}^4 \breve{\tau}^2 + O(\breve{\tau}^4)
\end{equation}
which matches the series expansion given for the bell sequence in Table \ref{tb:main_trans_results} up to a factor of 2, indicating that the $\tau^2$ scaling may be attributed to the sequence being open in transverse momentum.

The diamond sequence may be treated in a similar manner. Its transverse visibility may be approximated by the longitudinal visibility of the partially open sequence, Eq. \ref{eq:second_worst_lvis}. To leading order, the series expansion given by this approximation is:
\begin{equation}
    \mathcal{V}_{\perp}=|\mathcal{V}_{x}|^2 \approx 1-\left( \frac{32 \pi}{\grave{\lambda^2}}+\frac{4}{\grave{\sigma}_{\perp}^2} \right)\grave{\tau^4} + O(\grave{\tau}^6),
\end{equation}
where the length of the sequence has again been adjusted to $4\tau$. This expression may be rewritten as:
\begin{equation}
    \mathcal{V}_{\perp} \approx 1-\left( 4+\frac{32 \pi \breve{\sigma}_{\perp}^2}{\breve{\lambda^2}} \right)\breve{\tau^4} + O(\breve{\tau}^6),
\end{equation}
which differs from the series presented in Table \ref{tb:main_trans_results} by $4\breve{\tau}^4\ll1$. Similar to the bell sequence, this agreement indicates that the $\breve{\tau}^4$ scaling of the diamond sequence my be attributed to it being open in transverse position.

We conclude from this analogy that the different $\tau$ scaling of the transverse visibilities may be attributed to how ``open'' each sequence is. Within the linear approximation discussed in this section, the bell sequence is open in both position and momentum, the diamond sequence is open in position but closed in momentum, and the bow sequence is approximately closed in both. This progression corresponds to the observed improvements in transverse visibility discussed earlier in this section.
\markstopchange

\subsection{General considerations for other SGI systems}

\begin{figure}
\begin{center}
\includegraphics{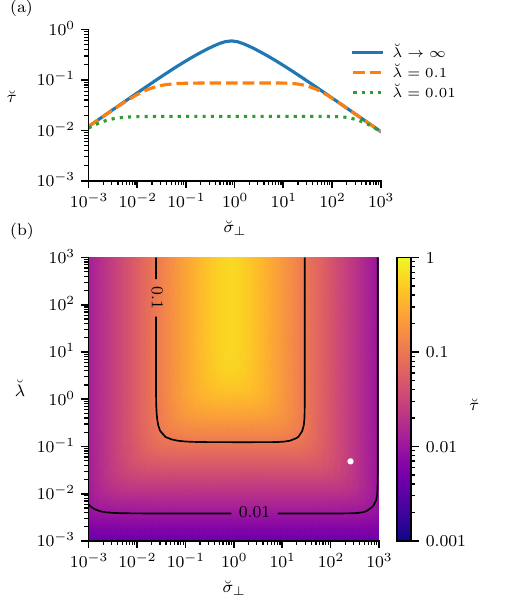}
\caption{\label{fg:tau_heat_map}
(a) Values of $\breve{\tau}$ corresponding to $\mathcal{V}_{\perp} =0.5$ for the bow sequence as a function of cloud size $\breve{\sigma}_{\perp}$.  Two different initial temperatures, corresponding to different $\breve{\lambda}$'s, are shown, together with the pure state case, $\breve{\lambda} \rightarrow \infty$.  (b) Similar to (a), but showing a continuous variation of $\breve{\lambda}$. The white spot corresponds to the proposed electric SGI (Table \ref{tb:ryd_summary}), with $\breve{\sigma}_{\perp} \approx 260$, $\breve{\lambda} \approx 0.048$, and $\breve{\tau}=0.023$ ($\tau \approx \SI{4.8}{\mu s}$) for $\mathcal{V}_{\perp} =0.5$. All calculations use the full expression for $\mathcal{V}_{\perp}$ \cite{github:our_repo}.
}
\end{center}
\end{figure}

Provided that the assumptions of our model are valid (linear energy shift with field strength, field variation given by Eq.~\ref{eq:e_field_variation_approx}), our expressions for transverse visibilities may be applied to other SGI systems.  Specifically, given the dimensionless parameters $\breve{\lambda}$, $\breve{\sigma}_{\perp}$, and $\breve{\tau}$, the transverse visibility is uniquely determined.  (See Eq.~\ref{eq:breve_def} and the surrounding discussion.)  Figure \ref{fg:tau_heat_map} illustrates the parameters required to obtain $\mathcal{V}_{\perp} = 0.5$ for the bow sequence.

Although not directly relevant to our proposed Rydberg SGI, Fig.~\ref{fg:tau_heat_map} shows it is possible for two values of $\breve{\sigma}_{\perp}$ to give the same $\mathcal{V}_{\perp}$ for a given $\breve{\lambda}$ and $\breve{\tau}$. These solutions are apparently logarithmically symmetric about $\breve{\sigma}_{\perp} = 2^{-1/4}$, which gives the highest attainable visibility. This value of $\breve{\sigma}_{\perp}$ represents a balance between the penalties for too large of a spread in the initial off-axis momenta (low $\breve{\sigma}_{\perp}$) versus positions (high $\breve{\sigma}_{\perp}$).

It is illustrative to consider the series expansion of $\mathcal{V}_{\perp}$ given in Table \ref{tb:main_trans_results} in the pure-state ($\breve{\lambda}\rightarrow\infty$) limit:
\begin{equation}
\label{eq:pure_bow_series}
\mathcal{V}_{\perp}=1 - \left(
-2
+ \dfrac{1}{2\breve{\sigma}_{\perp}^4}
+ 2 \breve{\sigma}_{\perp}^4
\right) \breve{\tau}^6
+  O(\breve{\tau}^8).
\end{equation}
Neglecting the higher order terms, this expression reaches a maximum at $\breve{\sigma}_{\perp} = 2^{-1/4}$. As the value of $\breve{\sigma}_{\perp}$ deviates from $2^{-1/4}$, the $\breve{\sigma}_{\perp}^4$ or $1/\breve{\sigma}_{\perp}^4$ portions of the $\breve{\tau}^6$ term dominate the loss in visibility, providing insight into the symmetry displayed in Fig.~\ref{fg:tau_heat_map}. Optimizing the full analytical expression for the visibility (in the pure state limit) also gives a maxima at $\breve{\sigma}_{\perp} = 2^{-1/4}$.

Figure \ref{fg:tau_heat_map}(a) also shows that there is a region of parameter space above the $\breve{\lambda}\rightarrow\infty$ curve where no solutions exist for $\mathcal{V}_{\perp}=0.5$. This region, in conjunction with the symmetry displayed in $\breve{\sigma}_{\perp}$, means that for a given $\breve{\tau}$ it is not always possible to achieve $\mathcal{V}_{\perp}=0.5$ by using an arbitrarily cold and/or small cloud.

\section{Discussion}
\label{se:discussion}

At first glance, all three SGI sequences that we have analyzed --- bell, diamond, and bow --- look plausible.  Under ideal conditions (considering only one spatial dimension) all of the sequences should have perfect visibilities; i.e., $\mathcal{V}_{\parallel}=1$.  However, as Comparat \cite{shortdoi:gptnmb} has shown, it is important to consider all three spatial dimensions.  We have considered a cylindrically symmetric geometry for the fields, and found that under certain simplifying assumptions the SGI fringe visibilities can be written as products $\mathcal{V} = \mathcal{V}_{\parallel} \mathcal{V}_{\perp}$, with the $\mathcal{V}_{\perp}$ factor arising from off-axis motion.  Analytical expressions can be obtained for this transverse visibility, $\mathcal{V}_{\perp}$, which indicate that as the duration $\tau$ of the interferometry sequences is increased, $\mathcal{V}_{\perp}$ for the bell sequence falls off more quickly than that of the diamond sequence, which in turn falls off more quickly than that of the bow sequence.

As a specific example of our general results, we analyzed a proposed electric SGI using Rb Rydberg atoms and found large practical differences between the sequences:  a transverse cloud width of \SI{1}{mm} is possible without appreciable loss in visibility for the bow sequence, whereas for the bell sequence the cloud width must be a factor of 1000 smaller for the same visibility. With a fixed maximum number density of Rydberg atoms, $n_{\text{Ryd}}$ --- chosen to avoid Rydberg-Rydberg interactions --- a bow sequence is much more experimentally feasible, since many more atoms can be used, lessening quantum projection noise \cite{shortdoi:c9q26q}; i.e., $N_{\text{Ryd}} \approx (2\pi)^{3/2} \: \sigma_{\perp}^2 \sigma_{\parallel} \: n_{\text{Ryd}}$.

Our analysis was based on: 1) a simplified model of how the field varies off-axis (Eq.~\ref{eq:e_field_variation_approx}), and 2) the assumption of a linear shift of internal state energies with field magnitude (Eq.~\ref{eq:linear_stark_shift}).  For the proposed electric SGI these are qualitatively reasonable assumptions ($d \gg \sigma_{\perp}$; see Table \ref{tb:ryd_summary} and Fig.~\ref{fg:stark_map}).   Nonetheless, due to the rich structure of the Rydberg system (see Fig.~\ref{fg:stark_map}), optimization of the visibility should consider non-linear energy shifts, such as done in the analysis of both Comparat \cite{shortdoi:gptnmb} and Zuniga \emph{et al.}~\cite{shortdoi:gt4jhh}. Also, the acceleration of the $\ket{g}$ state should be considered if it is comparable to that of the $\ket{r}$ state.  These extensions, including more accurate field models, could use analytical and numerical verification techniques similar to those presented in Ref.~\cite{shortdoi:gt63nj} in the context of photon-recoil matter-wave interferometry.

We have used the Rydberg atom SGI as an example --- our results are applicable to a broad class of SGIs that satisfy the same assumptions.  These include those using magnetic dipole moments and inhomogeneous magnetic fields as proposed for testing if gravity is quantum mechanical \cite{shortdoi:gcsb22, shortdoi:g94nnc_alt}.  And with slight modifications to our cylindrical field model, our approach is also applicable to ``planar'' geometries in which fields and field gradients are confined to a plane. Such geometries are suitable for SGI experiments that use atom chips, such as those demonstrated in Refs. \cite{shortdoi:gn7jtg, shortdoi:gkcvdb}.

\section*{Acknowledgements}

We thank F.~Luo for information concerning Rydberg state lifetimes,  H.~J.~Kim for field calculations, B.~Perez for useful discussions, and R.~Shiell for comments on this manuscript.


\bigbreak

\appendix

\markstartchange
\section{Initial state density matrix for a trapped harmonic oscillator potential}
\label{se:alt_dens_param}

With an appropriate choice of both $\lambda$ and $\sigma_x$, our initial state density matrix (Eq.~\ref{eq:combined_dens_pos}) can also model an initial state corresponding to atoms that --- just prior to the start of sequence --- were confined in a harmonic oscillator potential with temperature $T_{\text{HO}}$.  Here we provide this connection, discussing the one-dimensional case for simplicity, as it is easily extended to three-dimensions.

For the harmonic oscillator Hamiltonian $H = p^2/2m + m \omega^2 x^2 / 2$, let us define $l_0 \coloneqq \sqrt{\hbar/m\omega}$ and $\epsilon \coloneqq \hbar \omega / k T_{\text{HO}}$, adopting the notation of Ref.~\cite{shortdoi:g9wqxf} where it is shown that the associated density matrix is:
\begin{align} \label{eq:pos_dens_barragan}
\braket{x'|\,\hat{\rho}\,|x''} =&
\frac{1}{\sqrt{\pi l_0^2 \coth (\epsilon/2)}}
\times  \nonumber\\ 
&
\exp \bigg[
-\frac{1}{4l_0^2}
\Big\{
\coth \left( \frac{\epsilon}{2} \right) (x'-x'')^2 
\, + \nonumber \\
&
\qquad \tanh \left( \frac{\epsilon}{2} \right) (x'+x'')^2
\Big\}
\bigg].
\end{align}
An equivalent form involving hyperbolic trig functions of $\epsilon$ instead of $\epsilon/2$ is often presented \cite{isbn:9780201360769}, but the similarity of Eq.~\ref{eq:pos_dens_barragan} with Eq.~\ref{eq:density_matrix_with_sigma_p}, has the advantage that it allows us to quickly relate $\sigma_x$ and $\sigma_p$ to $l_0$ and $\epsilon$:
\begin{equation} \label{eq:sigmas_for_sho}
\sigma_x^2 = \frac{l_0^2}{2} \coth \frac{\epsilon}{2},
\quad
\text{and}
\quad
\sigma_p^2 = \frac{\hbar^2}{2 l_0^2} \coth \frac{\epsilon}{2}.
\end{equation}
Substitution of these expressions into Eq.~\ref{eq:lambda_given_sigmas} gives:
\begin{equation} \label{eq:lambda_for_tho}
\lambda = l_0 \sqrt{2\pi \sinh \epsilon},
\end{equation}
which along with the expression for $\sigma_x^2$ in Eq.~\ref{eq:sigmas_for_sho}
gives us the parameters that make the density matrix of Eq.~\ref{eq:combined_dens_pos} equivalent to that of Eq.~\ref{eq:pos_dens_barragan}.  In this way, our results for visibilities can be extended to physical systems for which the thermal harmonic oscillator model is more appropriate.

As an example of the correspondence between these two parameterizations, we find that for $\sigma = \SI{100}{\mu m}$ and $T = \SI{100}{\mu K}$ in the proposed Rydberg SGI (see Table \ref{tb:ryd_summary}), the equivalent trapped harmonic oscillator parameters are $l_0 \approx \SI{8.64e-7}{m} $ and $\epsilon \approx \SI{7.47e-5}{}$, which correspond to $\omega/2\pi \approx \SI{156}{Hz} $ and $T_{\text{HO}} \approx \SI{100}{\mu K}$.  (Since $|\epsilon| \ll 1$, then $\sinh \epsilon \approx \epsilon$ in Eq.~\ref{eq:lambda_for_tho}, which gives $T \approx T_{\text{HO}}$).

\markstopchange

\section{Rydberg state details}
\label{se:rydberg_details}

In the main text, we assessed the feasibility of an electric SGI with Rb Rydberg atoms of principal quantum number $n \approx 52$.
This $n$ was chosen because: 1) the electrode voltages for the required electric fields and gradients are reasonably small ($\approx \SI{10}{V}$, see Fig.~\ref{fg:electrodes}), 2) final state detection by selective field ionization does not require excessive electrode voltages ($\approx \SI{50}{V}$), and 3) the microwave transitions between low-angular momentum Rydberg states are at high enough frequencies to use waveguide horn antennas ($\approx \SI{10}{GHz}$) but are not so high as to require expensive sources.

Rydberg states decay by spontaneous emission and are also driven to other states by thermal radiation \cite{shortdoi:fctf3r}.  For $n \approx 50$ Rydberg atoms in a $T \approx \SI{300}{K}$ thermal radiation environment these loss mechanisms have similar contributions to a total loss rate of $R \approx \SI{e4}{s^{-1}}$ (see, for example, Ref.~\cite{shortdoi:b92t65}).  Thus, for a maximum loss of around $10\%$, the total time in Rydberg states should be limited to around $\approx \SI{10}{\mu s}$.  (The radiative lifetimes of the Rydberg atoms will be influenced by presence of the dc electric field $E_0 \approx \SI{210}{V/m}$. Mixing in higher angular momentum states will lengthen the zero dc-field spontaneous emission lifetimes, so our estimate of the total time available should be considered conservative.)

\begin{figure*}
\includegraphics{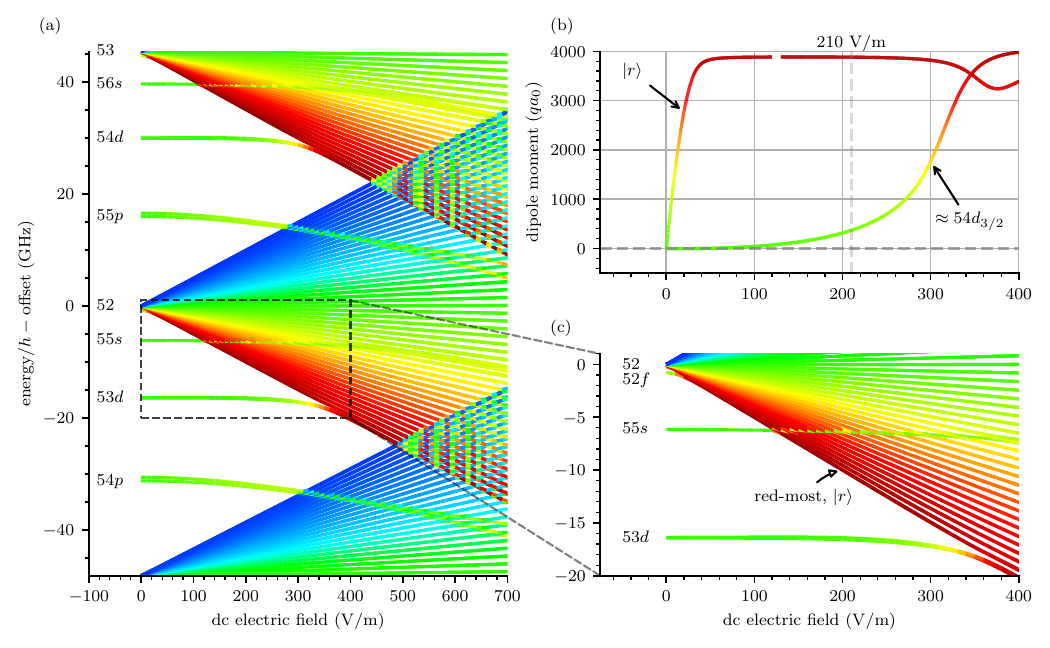}
\caption{
\label{fg:stark_map}
(a) and (c) Stark maps showing the energy levels of $m_j=1/2$ states of the Rb atom around $n=52$, with the colors indicating the dipole moment of the states. Computed using the techniques described in Ref.~\cite{shortdoi:dfv4zd} using the known zero field energies \cite{shortdoi:dntc3s, shortdoi:dc3462}.  The calculations include fine (but not hyperfine) structure.  Fine structure is barely resolvable in (a) and (c), so we exclude $j$ from the state labels. (b) Electric dipole moments of two states of interest (labeled) as a function of dc electric field (also computed using the techniques of Ref.~\cite{shortdoi:dfv4zd}; see main text).
}
\end{figure*}

Figure \ref{fg:stark_map}(a) shows the calculated Rb Rydberg energy levels around $n=52$ as a function of dc electric field --- a ``Stark map''.
We chose the $\ket{r}$ state to be the red-most, high-field seeking state of the $n=52$ manifold.  Based on its $f$-character, this state may be excited from the ground-state in a Doppler-free arrangement using a three-color laser system \cite{shortdoi:d5f2xr, shortdoi:dh5fxf}.

\markstartchange
As discussed in Section \ref{se:transverse_analysis}, for large $\mathcal{V}_{\perp}$ it is desirable to have the largest $E_0$ as possible, since by Eq.~\ref{eq:omega} this leads to an off-axis potential with lower $\omega$, and thus by Eq.~\ref{eq:breve_def}, a longer characteristic time $t_{\ebreve}$.  Although it is desirable to have as long as possible ``real'' times $\tau$ for the longitudinal motion, these are limited by the Rydberg lifetime. If we always use the maximum real time allowed by the Rydberg state, and thus hold $\tau$ fixed, a larger $t_{\ebreve}$ leads to a smaller $\breve{\tau}$, which gives improved $\mathcal{V}_{\perp}$ for all sequences (see Table \ref{tb:main_long_results}).

In addition to a large $E_0$, we also want the linear Stark shift model given by Eq.~\ref{eq:linear_stark_shift} to be as accurate as possible.  Based on these considerations, we choose $E_0 = \SI{210}{V/m}$, which is roughly centered between the avoided crossings of the $55s_{1/2}$ and $53d$ states with our choice of $\ket{r}$ (see Fig.~\ref{fg:stark_map}(c)).
\markstopchange

The same calculations used to create Figure \ref{fg:stark_map}(a) may also be used to determine the dipole moments $\mu_i$ that appear in Eq.~\ref{eq:linear_stark_shift} of the main text.  Specifically, the dipole moment $\mu_i$ of an energy eigenstate $i$ at a given electric field $E_0$ is determined by taking the expectation value of the same dipole matrix element operator used to compute the Stark map.  The magnitude of these dipole moments may be interpreted graphically as the magnitudes of the slopes of the adiabatically connected energies, with $\sgn \mu_j =1$ for a negative slope (high-field seeker) and $\sgn \mu_j =-1$ for a positive slope (low-field seeker).  For the $\ket{r}$ state at $E_0 = \SI{210}{V/m}$, $\mu_r \approx 3900 \: q a_0$, where $q$ is the elementary charge and $a_0$ is the Bohr radius.

The $\ket{g}$ state could be a ground electronic state with a negligible electric dipole moment in $E_0$.  But, it may be experimentally more convenient for $\ket{g}$ to be a Rydberg state. One choice is the state that is adiabatically connected with the zero field $54d_{3/2}$ state, which has a significantly lower dipole moment ($\mu_g \approx 370 \: q a_0$) than $\ket{r}$ at $E_0 = \SI{210}{V/m}$.

\section{Momentum space representations of the unitary time evolution and initial density operators}
\label{se:propagators}

In the main text, we make repeated use of the unitary time evolution operators in one spatial dimension for 1) a free particle, 2) a particle in a linear potential, and 3) a particle in a quadratic potential (inverted harmonic oscillator).  Since free particle evolution is diagonal in momentum space, it is convenient to express the initial density matrix in momentum space, and
all of the time-evolution operators as propagators in momentum space.  In this Appendix, we tabulate these well-known results from quantum-mechanics, omitting $x$, $y$, or $z$ component specifications for brevity.  In the main text, the density operator expressions and free particle propagators are used in both the longitudinal ($\uvect{z}$) and transverse ($\uvect{x}$ and $\uvect{y}$) directions, whereas the linear potential propagator is only used in the longitudinal direction ($\uvect{z}$).  The inverted harmonic oscillator propagator is only used for the transverse directions ($\uvect{x}$ and $\uvect{y}$).

\emph{Initial density operator in momentum space:}  A transformation of Eq.~\ref{eq:combined_dens_pos} to momentum space gives:
\begin{equation} \label{eq:appendix_init}
\braket{p_1|\hat{\rho}_{0,x}|p_2} = b \exp\left(
\sum_{j,k} B_{j,k} \: p_j p_k
\right)
\end{equation}
where
\begin{equation}
b = \frac{\sqrt{2} \lambda \sigma}{\hbar \sqrt{\pi} \sqrt{\lambda^2+8\pi\sigma^2}},
\end{equation}
and the summations in the exponent are over $j=1,2$ and $k=1,2$, with
\begin{equation}
B_{1,1} = B_{2,2} = - \frac{\sigma^2 (\lambda^2+4\pi\sigma^2)}{\hbar^2 (\lambda^2+8\pi\sigma^2)},
\end{equation}
and
\begin{equation} \label{eq:last_appendix_init}
B_{1,2} = B_{2,1} = \frac{4\pi\sigma^4}{\hbar^2 (\lambda^2+8\pi\sigma^2)}.
\end{equation}
The spatial width $\sigma$ should be replaced by $\sigma_{\parallel}$ when working in the longitudinal ($\uvect{z}$) direction and $\sigma_{\perp}$ when working in the transverse directions ($\uvect{x}$ and $\uvect{y}$).

Equation \ref{eq:appendix_init} may also be used to model a \emph{pure} initial state by taking the $\lambda \rightarrow \infty$ limit, resulting in $b=(2/\pi)^{1/2} \: \sigma / \hbar$, $B_{1,1} = B_{2,2} = -\sigma^2/\hbar^2$, and $B_{1,2}=B_{2,1}=0$.

\emph{Linear potential propagator:}
The effective potential for the atoms (Eq.~\ref{eq:eff_potential}) contains term linear in the longitudinal coordinate which may be condensed to $-F z$, where the corresponding force component is given by $F = 2 \mu_j E_0 \epsilon / d$.

Robinett \cite{shortdoi:cd9r2k} has presented a derivation of the propagator for a linear potential in momentum space (his Eq.~12).  With a potential of the form: $V = - F z$, this propagator may be written as:
\begin{multline} \label{eq:lin_prop_mom}
\braket{p_1|\hat{\mathcal{K}}_{\text{lin}}(t,F)|p_2} =
\delta(p_1-p_2-F t) \\ \times \exp\left(
-\frac{it}{2m\hbar}\left[
p_1^2-p_1 F t + \frac{1}{3}(F t)^2
\right]
\right).
\end{multline}
It is convenient to work in a unit system with the base units of length, momentum, and time defined as:
\begin{equation} \label{eq:check_def}
x_{\egrave} \coloneqq \left(\frac{\hbar^2}{m|F|} \right)^{1/3},
\quad
p_{\egrave} \coloneqq \frac{\hbar}{x_{\egrave}},
\quad
t_{\egrave} \coloneqq \frac{m}{\hbar} x_{\egrave}^2
\end{equation}
and use $\grave{\phantom{x}}$'s to denote quantities in this system; e.g., for an arbitrary length $\ell$ in the SI system, we have $\grave{\ell}=\ell/x_{\egrave}$. In this system, the propagator of Eq.~\ref{eq:lin_prop_mom} takes the form:
\begin{multline} \label{eq:prop_unit}
\braket{\grave{p}_1|\hat{\mathcal{K}}_{\text{lin}}(\grave{t},\grave{\Gamma})|\grave{p}_2} =
\delta(\grave{p}_1-\grave{p}_2 - \grave{\varGamma} \grave{t}) \\ \times \exp\left(
-\frac{i\grave{t}}{2}\left[
\grave{p}_1^2 - \grave{p}_1 \grave{\varGamma} \grave{t} + \frac{1}{3}\left(\grave{\varGamma} \grave{t} \right)^2
\right]
\right)
\end{multline}
where $\grave{\varGamma} = F/F_{\egrave}$. Frequently, the magnitude of a force used in a sequence is constant, but it may reverse direction (point towards $+\uvect{z}$ or $-\uvect{z}$).  In this case, the magnitude of the force defines the unit system through Eq.~\ref{eq:check_def}, and $\grave{\varGamma}$ corresponds to the direction of the force; i.e., $\grave{\varGamma} = 1$ for $+\uvect{z}$, and $\grave{\varGamma} = -1$ for $-\uvect{z}$.  Also, we condense the notation so that
$ \hat{\mathcal{K}}_{\text{lin}}(\grave{t},+) \equiv \hat{\mathcal{K}}_{\text{lin}}(\grave{t},+1)$
and
$ \hat{\mathcal{K}}_{\text{lin}}(\grave{t},-) \equiv \hat{\mathcal{K}}_{\text{lin}}(\grave{t},-1)$.

In the scaled unit system defined by Eq.~\ref{eq:check_def}, we have $\grave{\hbar} =1$. Thus, the initial state density matrix, as given by Eq.'s~\ref{eq:appendix_init} to \ref{eq:last_appendix_init},  can be expressed in this system by setting $\hbar=1$ and replacing $\sigma$ and $\lambda$ by $\grave{\sigma}$ and $\grave{\lambda}$.

\emph{Inverted harmonic oscillator propagator:}
The inverted harmonic oscillator Hamiltonian, $H = p^2/2m - m \omega^2 x^2 /2$, corresponds to the transverse terms $\uvect{x}$ and $\uvect{y}$ of the potential given by Eq.~\ref{eq:eff_potential}. It is convenient --- as customary with the non-inverted harmonic oscillator --- to work in a unit system with the base units of length, momentum, and time defined as:
\begin{equation} \label{eq:breve_def}
x_{\ebreve} \coloneqq \sqrt{\frac{\hbar}{m \omega}}
\quad
p_{\ebreve} \coloneqq \frac{\hbar}{x_{\ebreve}},
\quad
t_{\ebreve} \coloneqq \frac{1}{\omega}
\end{equation}
and use $\breve{\phantom{x}}$'s to denote quantities in this system; e.g., the Hamiltonian is now: $\breve{H} = \breve{p}^2/2 - \breve{x}^2/2$.  Examining Eq.~\ref{eq:eff_potential}, we find that
\begin{equation} \label{eq:omega}
\omega = \Big| \partial_z E_z|_0 \Big| \sqrt{\frac{\mu_r}{4mE_0}},
\end{equation}
indicating that increasing $E_0$ while keeping all other quantities constant results in a larger $t_{\ebreve}$, and thus smaller $\breve{\tau}$ for the same $\tau$.

Barton \cite{shortdoi:dnrjcn} gives the time-evolution operator in position space as (his Eq.~4.2):
\begin{multline}
\braket{\breve{x}_1|\hat{\mathcal{K}}_{\text{iho}}(\breve{t})|\breve{x}_2} =
\left[\frac{1}{2 \pi i \sinh \breve{t}}\right]^{1/2} \\ \times \exp \left\{\frac{i}{2 \sinh \breve{t}}\left[\left(\breve{x}^2_1+ \breve{x}^2_2 \right) \cosh \breve{t} -2 \breve{x}_1 \breve{x}_2 \right]\right\},
\end{multline}
which is obtained by the substitution of $\omega i$ for $\omega$ into the propagator for the normal (non-inverted) harmonic oscillator; e.g., Section 44 of Ref.~\cite{isbn:9780486414621}.  This propagator may be rewritten in momentum space as
\begin{equation}
\braket{\breve{p}_1|\hat{\mathcal{K}}_{\text{iho}}(\breve{t})|\breve{p}_2} = c \exp\left(
\sum_{j,k} C_{j,k} \: \breve{p}_j \breve{p}_k
\right)
\end{equation}
where $c = 1/(-2 \pi i \sinh \breve{t}\,)^{1/2}$ and
the summations in the exponent are over $j=1,2$ and $k=1,2$, with
\begin{equation}
C_{1,1} = C_{2,2} = - \frac{i}{2} \operatorname{coth} \breve{t}
\end{equation}
and
\begin{equation}
C_{1,2} = C_{2,1} = \frac{i}{2} \operatorname{csch} \breve{t}.
\end{equation}
To express the initial state density matrix (Eq.'s~\ref{eq:appendix_init} to \ref{eq:last_appendix_init}) in the harmonic oscillator unit system, set $\hbar=1$ and replace $\sigma$ and $\lambda$ by $\breve{\sigma}$ and $\breve{\lambda}$.

\emph{Free particle propagator:}
When using the free particle propagator
\begin{equation}
\braket{p_1|\hat{\mathcal{K}}_{\text{free}}(t)|p_2} =
\delta(p_1-p_2)
\exp\left(
-i \: \frac{p_1^2}{2m}  \frac{t}{\hbar} \right)
\end{equation}
to analyze longitudinal motion, it is convenient to express it in the same unit system as Eq.~\ref{eq:prop_unit}; i.e.,
\begin{equation}
\braket{\grave{p}_1|\hat{\mathcal{K}}_{\text{free}}(\grave{t})|\grave{p}_2} =
\delta(\grave{p}_1-\grave{p}_2)
\exp\left(
-\frac{i}{2} \: \grave{p}_1^2  t \right).
\end{equation}
Likewise, when considering transverse motion, the free particle propagator has a similar form when expressed in the appropriate unit system (replace $\grave{\phantom{x}}$'s with $\breve{\phantom{x}}$'s).\\

\section{Explicit expression for the transverse visibility of the bell sequence}
\label{se:vis_bell}

As discussed in the main text, the exact expressions for the transverse visibilities are quite lengthy.  For the bell sequence, we find that:
\begin{widetext}
\begin{multline} \label{eq:full_bell_expression}
\mathcal{V}_{\perp} = \frac{4 \breve{\lambda} ^2 \breve{\sigma}_{\perp} ^2}{\sqrt{16 \breve{\sigma}_{\perp} ^4 \left[\left\{\breve{\lambda} ^2+4 \pi  \breve{\sigma}_{\perp} ^2\right\} \{\operatorname{C} (4 \breve{\tau} )-2 \breve{\tau}  \operatorname{S} (4 \breve{\tau} )\} -4 \pi  \breve{\sigma}_{\perp} ^2\right]^2+\left[4 \breve{\tau}  \left\{\breve{\lambda} ^2+8 \pi  \breve{\sigma}_{\perp} ^2\right\} \operatorname{C} (4 \breve{\tau} )+\left\{\breve{\lambda} ^2 \left(4 \breve{\sigma}_{\perp} ^4-1\right)-8 \pi  \breve{\sigma}_{\perp} ^2\right\} \operatorname{S} (4 \breve{\tau} )\right]^2}}
\end{multline}
\end{widetext}
where $\operatorname{S}$ and $\operatorname{C}$ are the hyperbolic trigonometric functions $\sinh$ and $\cosh$.

Replacing $\breve{\tau}$ by $\breve{\tau}/4$ in Eq.~\ref{eq:full_bell_expression}, gives the transverse visibility $\mathcal{V}_{\perp}$ for the fully-open sequence discussed in Section \ref{se:destruction_by_sg}.  Likewise, replacing $\breve{\tau}$ by $\breve{\tau}/2$ in the full expression for the diamond sequence
\cite{github:our_repo} gives the $\mathcal{V}_{\perp}$ corresponding to the partially open sequence discussed in Section \ref{se:measurement_of_lambda_t}.

The exact expressions for the diamond and bow sequences are significantly more complicated than Eq.~\ref{eq:full_bell_expression} but may be reproduced using the code found in Ref.~\cite{github:our_repo} (which also contains \textsc{Python} code for their numerical evaluation).

\newpage


\begin{thebibliography}{61}%
\makeatletter
\providecommand \@ifxundefined [1]{%
 \@ifx{#1\undefined}
}%
\providecommand \@ifnum [1]{%
 \ifnum #1\expandafter \@firstoftwo
 \else \expandafter \@secondoftwo
 \fi
}%
\providecommand \@ifx [1]{%
 \ifx #1\expandafter \@firstoftwo
 \else \expandafter \@secondoftwo
 \fi
}%
\providecommand \natexlab [1]{#1}%
\providecommand \enquote  [1]{``#1''}%
\providecommand \bibnamefont  [1]{#1}%
\providecommand \bibfnamefont [1]{#1}%
\providecommand \citenamefont [1]{#1}%
\providecommand \href@noop [0]{\@secondoftwo}%
\providecommand \href [0]{\begingroup \@sanitize@url \@href}%
\providecommand \@href[1]{\@@startlink{#1}\@@href}%
\providecommand \@@href[1]{\endgroup#1\@@endlink}%
\providecommand \@sanitize@url [0]{\catcode `\\12\catcode `\$12\catcode
  `\&12\catcode `\#12\catcode `\^12\catcode `\_12\catcode `\%12\relax}%
\providecommand \@@startlink[1]{}%
\providecommand \@@endlink[0]{}%
\providecommand \url  [0]{\begingroup\@sanitize@url \@url }%
\providecommand \@url [1]{\endgroup\@href {#1}{\urlprefix }}%
\providecommand \urlprefix  [0]{URL }%
\providecommand \Eprint [0]{\href }%
\providecommand \doibase [0]{http://dx.doi.org/}%
\providecommand \selectlanguage [0]{\@gobble}%
\providecommand \bibinfo  [0]{\@secondoftwo}%
\providecommand \bibfield  [0]{\@secondoftwo}%
\providecommand \translation [1]{[#1]}%
\providecommand \BibitemOpen [0]{}%
\providecommand \bibitemStop [0]{}%
\providecommand \bibitemNoStop [0]{.\EOS\space}%
\providecommand \EOS [0]{\spacefactor3000\relax}%
\providecommand \BibitemShut  [1]{\csname bibitem#1\endcsname}%
\let\auto@bib@innerbib\@empty
\bibitem [{\citenamefont {Opie}\ \emph {et~al.}(2007)\citenamefont {Opie},
  \citenamefont {Opie},\ and\ \citenamefont {Hassall}}]{isbn:9780198691129}%
  \BibitemOpen
  \bibinfo {editor} {\bibfnamefont {I.}~\bibnamefont {Opie}}, \bibinfo {editor}
  {\bibfnamefont {P.}~\bibnamefont {Opie}}, \ and\ \bibinfo {editor}
  {\bibfnamefont {J.}~\bibnamefont {Hassall}},\ eds.,\ \href@noop {} {\emph
  {\bibinfo {title} {The {{Oxford}} Nursery Rhyme Book}}},\ \bibinfo {edition}
  {repr.}\ ed.\ (\bibinfo  {publisher} {Oxford University Press},\ \bibinfo
  {address} {Oxford},\ \bibinfo {year} {2007})\BibitemShut {NoStop}%
\bibitem [{\citenamefont {Englert}\ \emph {et~al.}(1988)\citenamefont
  {Englert}, \citenamefont {Schwinger},\ and\ \citenamefont
  {Scully}}]{shortdoi:c34q7x}%
  \BibitemOpen
  \bibfield  {author} {\bibinfo {author} {\bibfnamefont {B.-G.}\ \bibnamefont
  {Englert}}, \bibinfo {author} {\bibfnamefont {J.}~\bibnamefont {Schwinger}},
  \ and\ \bibinfo {author} {\bibfnamefont {M.~O.}\ \bibnamefont {Scully}},\
  }\href {\doibase 10.1007/BF01909939} {\bibfield  {journal} {\bibinfo
  {journal} {Foundations of Physics}\ }\textbf {\bibinfo {volume} {18}},\
  \bibinfo {pages} {1045} (\bibinfo {year} {1988})}\BibitemShut {NoStop}%
\bibitem [{\citenamefont {Bohm}(1989)}]{isbn:9780486659695}%
  \BibitemOpen
  \bibfield  {author} {\bibinfo {author} {\bibfnamefont {D.}~\bibnamefont
  {Bohm}},\ }\href@noop {} {\emph {\bibinfo {title} {Quantum Theory}}}\
  (\bibinfo  {publisher} {Dover Publications},\ \bibinfo {address} {New York},\
  \bibinfo {year} {1989})\BibitemShut {NoStop}%
\bibitem [{\citenamefont {Wigner}(1963)}]{shortdoi:bdgdmp}%
  \BibitemOpen
  \bibfield  {author} {\bibinfo {author} {\bibfnamefont {E.~P.}\ \bibnamefont
  {Wigner}},\ }\href {\doibase 10.1119/1.1969254} {\bibfield  {journal}
  {\bibinfo  {journal} {American Journal of Physics}\ }\textbf {\bibinfo
  {volume} {31}},\ \bibinfo {pages} {6} (\bibinfo {year} {1963})}\BibitemShut
  {NoStop}%
\bibitem [{\citenamefont {Amit}\ \emph {et~al.}(2019)\citenamefont {Amit},
  \citenamefont {Margalit}, \citenamefont {Dobkowski}, \citenamefont {Zhou},
  \citenamefont {Japha}, \citenamefont {Zimmermann}, \citenamefont {Efremov},
  \citenamefont {Narducci}, \citenamefont {Rasel}, \citenamefont {Schleich},\
  and\ \citenamefont {Folman}}]{shortdoi:gn7jtg}%
  \BibitemOpen
  \bibfield  {author} {\bibinfo {author} {\bibfnamefont {O.}~\bibnamefont
  {Amit}}, \bibinfo {author} {\bibfnamefont {Y.}~\bibnamefont {Margalit}},
  \bibinfo {author} {\bibfnamefont {O.}~\bibnamefont {Dobkowski}}, \bibinfo
  {author} {\bibfnamefont {Z.}~\bibnamefont {Zhou}}, \bibinfo {author}
  {\bibfnamefont {Y.}~\bibnamefont {Japha}}, \bibinfo {author} {\bibfnamefont
  {M.}~\bibnamefont {Zimmermann}}, \bibinfo {author} {\bibfnamefont {M.~A.}\
  \bibnamefont {Efremov}}, \bibinfo {author} {\bibfnamefont {F.~A.}\
  \bibnamefont {Narducci}}, \bibinfo {author} {\bibfnamefont {E.~M.}\
  \bibnamefont {Rasel}}, \bibinfo {author} {\bibfnamefont {W.~P.}\ \bibnamefont
  {Schleich}}, \ and\ \bibinfo {author} {\bibfnamefont {R.}~\bibnamefont
  {Folman}},\ }\href {\doibase 10.1103/PhysRevLett.123.083601} {\bibfield
  {journal} {\bibinfo  {journal} {Physical Review Letters}\ }\textbf {\bibinfo
  {volume} {123}},\ \bibinfo {pages} {083601} (\bibinfo {year}
  {2019})}\BibitemShut {NoStop}%
\bibitem [{\citenamefont {Margalit}\ \emph {et~al.}(2021)\citenamefont
  {Margalit}, \citenamefont {Dobkowski}, \citenamefont {Zhou}, \citenamefont
  {Amit}, \citenamefont {Japha}, \citenamefont {Moukouri}, \citenamefont
  {Rohrlich}, \citenamefont {Mazumdar}, \citenamefont {Bose}, \citenamefont
  {Henkel},\ and\ \citenamefont {Folman}}]{shortdoi:gkcvdb}%
  \BibitemOpen
  \bibfield  {author} {\bibinfo {author} {\bibfnamefont {Y.}~\bibnamefont
  {Margalit}}, \bibinfo {author} {\bibfnamefont {O.}~\bibnamefont {Dobkowski}},
  \bibinfo {author} {\bibfnamefont {Z.}~\bibnamefont {Zhou}}, \bibinfo {author}
  {\bibfnamefont {O.}~\bibnamefont {Amit}}, \bibinfo {author} {\bibfnamefont
  {Y.}~\bibnamefont {Japha}}, \bibinfo {author} {\bibfnamefont
  {S.}~\bibnamefont {Moukouri}}, \bibinfo {author} {\bibfnamefont
  {D.}~\bibnamefont {Rohrlich}}, \bibinfo {author} {\bibfnamefont
  {A.}~\bibnamefont {Mazumdar}}, \bibinfo {author} {\bibfnamefont
  {S.}~\bibnamefont {Bose}}, \bibinfo {author} {\bibfnamefont {C.}~\bibnamefont
  {Henkel}}, \ and\ \bibinfo {author} {\bibfnamefont {R.}~\bibnamefont
  {Folman}},\ }\href {\doibase 10.1126/sciadv.abg2879} {\bibfield  {journal}
  {\bibinfo  {journal} {Science Advances}\ }\textbf {\bibinfo {volume} {7}},\
  \bibinfo {pages} {eabg2879} (\bibinfo {year} {2021})}\BibitemShut {NoStop}%
\bibitem [{\citenamefont {Peters}\ \emph {et~al.}(2001)\citenamefont {Peters},
  \citenamefont {Chung},\ and\ \citenamefont {Chu}}]{shortdoi:fdtkbw}%
  \BibitemOpen
  \bibfield  {author} {\bibinfo {author} {\bibfnamefont {A.}~\bibnamefont
  {Peters}}, \bibinfo {author} {\bibfnamefont {K.~Y.}\ \bibnamefont {Chung}}, \
  and\ \bibinfo {author} {\bibfnamefont {S.}~\bibnamefont {Chu}},\ }\href
  {\doibase 10.1088/0026-1394/38/1/4} {\bibfield  {journal} {\bibinfo
  {journal} {Metrologia}\ }\textbf {\bibinfo {volume} {38}},\ \bibinfo {pages}
  {25} (\bibinfo {year} {2001})}\BibitemShut {NoStop}%
\bibitem [{\citenamefont {Bongs}\ \emph {et~al.}(2019)\citenamefont {Bongs},
  \citenamefont {Holynski}, \citenamefont {Vovrosh}, \citenamefont {Bouyer},
  \citenamefont {Condon}, \citenamefont {Rasel}, \citenamefont {Schubert},
  \citenamefont {Schleich},\ and\ \citenamefont {Roura}}]{shortdoi:ggxx4v}%
  \BibitemOpen
  \bibfield  {author} {\bibinfo {author} {\bibfnamefont {K.}~\bibnamefont
  {Bongs}}, \bibinfo {author} {\bibfnamefont {M.}~\bibnamefont {Holynski}},
  \bibinfo {author} {\bibfnamefont {J.}~\bibnamefont {Vovrosh}}, \bibinfo
  {author} {\bibfnamefont {P.}~\bibnamefont {Bouyer}}, \bibinfo {author}
  {\bibfnamefont {G.}~\bibnamefont {Condon}}, \bibinfo {author} {\bibfnamefont
  {E.}~\bibnamefont {Rasel}}, \bibinfo {author} {\bibfnamefont
  {C.}~\bibnamefont {Schubert}}, \bibinfo {author} {\bibfnamefont {W.~P.}\
  \bibnamefont {Schleich}}, \ and\ \bibinfo {author} {\bibfnamefont
  {A.}~\bibnamefont {Roura}},\ }\href {\doibase 10.1038/s42254-019-0117-4}
  {\bibfield  {journal} {\bibinfo  {journal} {Nature Reviews Physics}\ }\textbf
  {\bibinfo {volume} {1}},\ \bibinfo {pages} {731} (\bibinfo {year}
  {2019})}\BibitemShut {NoStop}%
\bibitem [{\citenamefont {Zimmermann}\ \emph {et~al.}(2017)\citenamefont
  {Zimmermann}, \citenamefont {Efremov}, \citenamefont {Roura}, \citenamefont
  {Schleich}, \citenamefont {DeSavage}, \citenamefont {Davis}, \citenamefont
  {Srinivasan}, \citenamefont {Narducci}, \citenamefont {Werner},\ and\
  \citenamefont {Rasel}}]{shortdoi:gpb7cs}%
  \BibitemOpen
  \bibfield  {author} {\bibinfo {author} {\bibfnamefont {M.}~\bibnamefont
  {Zimmermann}}, \bibinfo {author} {\bibfnamefont {M.~A.}\ \bibnamefont
  {Efremov}}, \bibinfo {author} {\bibfnamefont {A.}~\bibnamefont {Roura}},
  \bibinfo {author} {\bibfnamefont {W.~P.}\ \bibnamefont {Schleich}}, \bibinfo
  {author} {\bibfnamefont {S.~A.}\ \bibnamefont {DeSavage}}, \bibinfo {author}
  {\bibfnamefont {J.~P.}\ \bibnamefont {Davis}}, \bibinfo {author}
  {\bibfnamefont {A.}~\bibnamefont {Srinivasan}}, \bibinfo {author}
  {\bibfnamefont {F.~A.}\ \bibnamefont {Narducci}}, \bibinfo {author}
  {\bibfnamefont {S.~A.}\ \bibnamefont {Werner}}, \ and\ \bibinfo {author}
  {\bibfnamefont {E.~M.}\ \bibnamefont {Rasel}},\ }\href {\doibase
  10.1007/s00340-017-6655-5} {\bibfield  {journal} {\bibinfo  {journal}
  {Applied Physics B}\ }\textbf {\bibinfo {volume} {123}},\ \bibinfo {pages}
  {102} (\bibinfo {year} {2017})}\BibitemShut {NoStop}%
\bibitem [{\citenamefont {Comparat}(2020)}]{shortdoi:gptnmb}%
  \BibitemOpen
  \bibfield  {author} {\bibinfo {author} {\bibfnamefont {D.}~\bibnamefont
  {Comparat}},\ }\href {\doibase 10.1103/PhysRevA.101.023606} {\bibfield
  {journal} {\bibinfo  {journal} {Physical Review A}\ }\textbf {\bibinfo
  {volume} {101}},\ \bibinfo {pages} {023606} (\bibinfo {year}
  {2020})}\BibitemShut {NoStop}%
\bibitem [{\citenamefont {Zuniga}\ \emph {et~al.}(2024)\citenamefont {Zuniga},
  \citenamefont {Gomez},\ and\ \citenamefont
  {{Castanos-Cervantes}}}]{shortdoi:gt4jhh}%
  \BibitemOpen
  \bibfield  {author} {\bibinfo {author} {\bibfnamefont {E.}~\bibnamefont
  {Zuniga}}, \bibinfo {author} {\bibfnamefont {E.}~\bibnamefont {Gomez}}, \
  and\ \bibinfo {author} {\bibfnamefont {L.~O.}\ \bibnamefont
  {{Castanos-Cervantes}}},\ }\href {\doibase 10.1103/PhysRevA.109.013304}
  {\bibfield  {journal} {\bibinfo  {journal} {Physical Review A}\ }\textbf
  {\bibinfo {volume} {109}},\ \bibinfo {pages} {013304} (\bibinfo {year}
  {2024})}\BibitemShut {NoStop}%
\bibitem [{\citenamefont {Bose}\ \emph {et~al.}(2017)\citenamefont {Bose},
  \citenamefont {Mazumdar}, \citenamefont {Morley}, \citenamefont {Ulbricht},
  \citenamefont {Toro{\v s}}, \citenamefont {Paternostro}, \citenamefont
  {Geraci}, \citenamefont {Barker}, \citenamefont {Kim},\ and\ \citenamefont
  {Milburn}}]{shortdoi:gcsb22}%
  \BibitemOpen
  \bibfield  {author} {\bibinfo {author} {\bibfnamefont {S.}~\bibnamefont
  {Bose}}, \bibinfo {author} {\bibfnamefont {A.}~\bibnamefont {Mazumdar}},
  \bibinfo {author} {\bibfnamefont {G.~W.}\ \bibnamefont {Morley}}, \bibinfo
  {author} {\bibfnamefont {H.}~\bibnamefont {Ulbricht}}, \bibinfo {author}
  {\bibfnamefont {M.}~\bibnamefont {Toro{\v s}}}, \bibinfo {author}
  {\bibfnamefont {M.}~\bibnamefont {Paternostro}}, \bibinfo {author}
  {\bibfnamefont {A.~A.}\ \bibnamefont {Geraci}}, \bibinfo {author}
  {\bibfnamefont {P.~F.}\ \bibnamefont {Barker}}, \bibinfo {author}
  {\bibfnamefont {M.~S.}\ \bibnamefont {Kim}}, \ and\ \bibinfo {author}
  {\bibfnamefont {G.}~\bibnamefont {Milburn}},\ }\href {\doibase
  10.1103/PhysRevLett.119.240401} {\bibfield  {journal} {\bibinfo  {journal}
  {Physical Review Letters}\ }\textbf {\bibinfo {volume} {119}},\ \bibinfo
  {pages} {240401} (\bibinfo {year} {2017})}\BibitemShut {NoStop}%
\bibitem [{\citenamefont {Bose}\ \emph {et~al.}(2025)\citenamefont {Bose} \emph
  {et~al.}}]{shortdoi:g94nnc_alt}%
  \BibitemOpen
  \bibfield  {author} {\bibinfo {author} {\bibfnamefont {S.}~\bibnamefont
  {Bose}} \emph {et~al.},\ }\href {\doibase 10.48550/arXiv.2509.01586}
  {\enquote {\bibinfo {title} {A {{Spin-Based Pathway}} to {{Testing}} the
  {{Quantum Nature}} of {{Gravity}}},}\ } (\bibinfo {year} {2025}),\ \Eprint
  {http://arxiv.org/abs/2509.01586} {arXiv:2509.01586} \BibitemShut {NoStop}%
\bibitem [{\citenamefont {Oh}\ \emph {et~al.}(2020)\citenamefont {Oh},
  \citenamefont {Kwon}, \citenamefont {Jiang},\ and\ \citenamefont
  {Kim}}]{shortdoi:g9hccm}%
  \BibitemOpen
  \bibfield  {author} {\bibinfo {author} {\bibfnamefont {C.}~\bibnamefont
  {Oh}}, \bibinfo {author} {\bibfnamefont {H.}~\bibnamefont {Kwon}}, \bibinfo
  {author} {\bibfnamefont {L.}~\bibnamefont {Jiang}}, \ and\ \bibinfo {author}
  {\bibfnamefont {M.~S.}\ \bibnamefont {Kim}},\ }\href {\doibase
  10.1103/PhysRevA.102.053321} {\bibfield  {journal} {\bibinfo  {journal}
  {Physical Review A}\ }\textbf {\bibinfo {volume} {102}},\ \bibinfo {pages}
  {053321} (\bibinfo {year} {2020})}\BibitemShut {NoStop}%
\bibitem [{\citenamefont {Hajebrahimi}\ \emph {et~al.}(2023)\citenamefont
  {Hajebrahimi}, \citenamefont {Manshouri}, \citenamefont {Sharifian},\ and\
  \citenamefont {Zarei}}]{shortdoi:hbdsg9}%
  \BibitemOpen
  \bibfield  {author} {\bibinfo {author} {\bibfnamefont {M.}~\bibnamefont
  {Hajebrahimi}}, \bibinfo {author} {\bibfnamefont {H.}~\bibnamefont
  {Manshouri}}, \bibinfo {author} {\bibfnamefont {M.}~\bibnamefont
  {Sharifian}}, \ and\ \bibinfo {author} {\bibfnamefont {M.}~\bibnamefont
  {Zarei}},\ }\href {\doibase 10.1140/epjc/s10052-022-11152-9} {\bibfield
  {journal} {\bibinfo  {journal} {The European Physical Journal C}\ }\textbf
  {\bibinfo {volume} {83}},\ \bibinfo {pages} {11} (\bibinfo {year}
  {2023})}\BibitemShut {NoStop}%
\bibitem [{\citenamefont {Xiang}\ \emph {et~al.}(2024)\citenamefont {Xiang},
  \citenamefont {Zhou}, \citenamefont {Bose},\ and\ \citenamefont
  {Mazumdar}}]{xiang_phonon_induced_2024}%
  \BibitemOpen
  \bibfield  {author} {\bibinfo {author} {\bibfnamefont {Q.}~\bibnamefont
  {Xiang}}, \bibinfo {author} {\bibfnamefont {R.}~\bibnamefont {Zhou}},
  \bibinfo {author} {\bibfnamefont {S.}~\bibnamefont {Bose}}, \ and\ \bibinfo
  {author} {\bibfnamefont {A.}~\bibnamefont {Mazumdar}},\ }\href {\doibase
  10.1103/PhysRevA.110.042614} {\bibfield  {journal} {\bibinfo  {journal}
  {Physical Review A}\ }\textbf {\bibinfo {volume} {110}},\ \bibinfo {pages}
  {042614} (\bibinfo {year} {2024})}\BibitemShut {NoStop}%
\bibitem [{\citenamefont {Henkel}\ and\ \citenamefont
  {Folman}(2022)}]{henkel_internal_2022}%
  \BibitemOpen
  \bibfield  {author} {\bibinfo {author} {\bibfnamefont {C.}~\bibnamefont
  {Henkel}}\ and\ \bibinfo {author} {\bibfnamefont {R.}~\bibnamefont
  {Folman}},\ }\href {\doibase 10.1116/5.0080503} {\bibfield  {journal}
  {\bibinfo  {journal} {AVS Quantum Science}\ }\textbf {\bibinfo {volume}
  {4}},\ \bibinfo {pages} {025602} (\bibinfo {year} {2022})}\BibitemShut
  {NoStop}%
\bibitem [{\citenamefont {Henkel}\ and\ \citenamefont
  {Folman}(2024)}]{henkel_universal_2024}%
  \BibitemOpen
  \bibfield  {author} {\bibinfo {author} {\bibfnamefont {C.}~\bibnamefont
  {Henkel}}\ and\ \bibinfo {author} {\bibfnamefont {R.}~\bibnamefont
  {Folman}},\ }\href {\doibase 10.1103/PhysRevA.110.042221} {\bibfield
  {journal} {\bibinfo  {journal} {Physical Review A}\ }\textbf {\bibinfo
  {volume} {110}},\ \bibinfo {pages} {042221} (\bibinfo {year}
  {2024})}\BibitemShut {NoStop}%
\bibitem [{\citenamefont {Wu}\ \emph {et~al.}(2025)\citenamefont {Wu},
  \citenamefont {Toroš}, \citenamefont {Bose},\ and\ \citenamefont
  {Mazumdar}}]{wu_inertial_2025}%
  \BibitemOpen
  \bibfield  {author} {\bibinfo {author} {\bibfnamefont {M.-Z.}\ \bibnamefont
  {Wu}}, \bibinfo {author} {\bibfnamefont {M.}~\bibnamefont {Toroš}}, \bibinfo
  {author} {\bibfnamefont {S.}~\bibnamefont {Bose}}, \ and\ \bibinfo {author}
  {\bibfnamefont {A.}~\bibnamefont {Mazumdar}},\ }\href {\doibase
  10.1103/PhysRevD.111.064004} {\bibfield  {journal} {\bibinfo  {journal}
  {Physical Review D}\ }\textbf {\bibinfo {volume} {111}},\ \bibinfo {pages}
  {064004} (\bibinfo {year} {2025})}\BibitemShut {NoStop}%
\bibitem [{\citenamefont {Henkel}\ \emph {et~al.}(2003)\citenamefont {Henkel},
  \citenamefont {Krüger}, \citenamefont {Folman},\ and\ \citenamefont
  {Schmiedmayer}}]{henkel_fundamental_2003}%
  \BibitemOpen
  \bibfield  {author} {\bibinfo {author} {\bibfnamefont {C.}~\bibnamefont
  {Henkel}}, \bibinfo {author} {\bibfnamefont {P.}~\bibnamefont {Krüger}},
  \bibinfo {author} {\bibfnamefont {R.}~\bibnamefont {Folman}}, \ and\ \bibinfo
  {author} {\bibfnamefont {J.}~\bibnamefont {Schmiedmayer}},\ }\href {\doibase
  10.1007/s00340-003-1112-z} {\bibfield  {journal} {\bibinfo  {journal}
  {Applied Physics B}\ }\textbf {\bibinfo {volume} {76}},\ \bibinfo {pages}
  {173} (\bibinfo {year} {2003})}\BibitemShut {NoStop}%
\bibitem [{\citenamefont {Narasimha~Moorthy}\ and\ \citenamefont
  {Mazumdar}(2026)}]{narasimha_moorthy_magnetic_2026}%
  \BibitemOpen
  \bibfield  {author} {\bibinfo {author} {\bibfnamefont {S.}~\bibnamefont
  {Narasimha~Moorthy}}\ and\ \bibinfo {author} {\bibfnamefont {A.}~\bibnamefont
  {Mazumdar}},\ }\href {\doibase 10.1103/nkr5-9dxr} {\bibfield  {journal}
  {\bibinfo  {journal} {Physical Review A}\ }\textbf {\bibinfo {volume}
  {113}},\ \bibinfo {pages} {012419} (\bibinfo {year} {2026})}\BibitemShut
  {NoStop}%
\bibitem [{\citenamefont {Zhou}\ \emph
  {et~al.}(2025{\natexlab{a}})\citenamefont {Zhou}, \citenamefont {Rizaldy},
  \citenamefont {Schut},\ and\ \citenamefont {Mazumdar}}]{zhou_spin_2025}%
  \BibitemOpen
  \bibfield  {author} {\bibinfo {author} {\bibfnamefont {T.}~\bibnamefont
  {Zhou}}, \bibinfo {author} {\bibfnamefont {R.}~\bibnamefont {Rizaldy}},
  \bibinfo {author} {\bibfnamefont {M.}~\bibnamefont {Schut}}, \ and\ \bibinfo
  {author} {\bibfnamefont {A.}~\bibnamefont {Mazumdar}},\ }\href {\doibase
  10.1103/dbrs-wn92} {\bibfield  {journal} {\bibinfo  {journal} {Physical
  Review A}\ }\textbf {\bibinfo {volume} {112}},\ \bibinfo {pages} {012613}
  (\bibinfo {year} {2025}{\natexlab{a}})}\BibitemShut {NoStop}%
\bibitem [{\citenamefont {Zhou}\ \emph
  {et~al.}(2025{\natexlab{b}})\citenamefont {Zhou}, \citenamefont {Bose},\ and\
  \citenamefont {Mazumdar}}]{zhou_gyroscopic_2025}%
  \BibitemOpen
  \bibfield  {author} {\bibinfo {author} {\bibfnamefont {T.}~\bibnamefont
  {Zhou}}, \bibinfo {author} {\bibfnamefont {S.}~\bibnamefont {Bose}}, \ and\
  \bibinfo {author} {\bibfnamefont {A.}~\bibnamefont {Mazumdar}},\ }\href
  {\doibase 10.1103/4xvz-gnk7} {\bibfield  {journal} {\bibinfo  {journal}
  {Physical Review A}\ }\textbf {\bibinfo {volume} {112}},\ \bibinfo {pages}
  {013315} (\bibinfo {year} {2025}{\natexlab{b}})}\BibitemShut {NoStop}%
\bibitem [{\citenamefont {Japha}\ and\ \citenamefont
  {Folman}(2023)}]{japha_quantum_2023}%
  \BibitemOpen
  \bibfield  {author} {\bibinfo {author} {\bibfnamefont {Y.}~\bibnamefont
  {Japha}}\ and\ \bibinfo {author} {\bibfnamefont {R.}~\bibnamefont {Folman}},\
  }\href {\doibase 10.1103/PhysRevLett.130.113602} {\bibfield  {journal}
  {\bibinfo  {journal} {Physical Review Letters}\ }\textbf {\bibinfo {volume}
  {130}},\ \bibinfo {pages} {113602} (\bibinfo {year} {2023})}\BibitemShut
  {NoStop}%
\bibitem [{\citenamefont {Rizaldy}\ \emph {et~al.}(2025)\citenamefont
  {Rizaldy}, \citenamefont {Zhou}, \citenamefont {Bose},\ and\ \citenamefont
  {Mazumdar}}]{rizaldy_rotational_2025}%
  \BibitemOpen
  \bibfield  {author} {\bibinfo {author} {\bibfnamefont {R.}~\bibnamefont
  {Rizaldy}}, \bibinfo {author} {\bibfnamefont {T.}~\bibnamefont {Zhou}},
  \bibinfo {author} {\bibfnamefont {S.}~\bibnamefont {Bose}}, \ and\ \bibinfo
  {author} {\bibfnamefont {A.}~\bibnamefont {Mazumdar}},\ }\href {\doibase
  10.1103/vjms-5bqd} {\bibfield  {journal} {\bibinfo  {journal} {Physical
  Review Research}\ }\textbf {\bibinfo {volume} {7}},\ \bibinfo {pages}
  {043095} (\bibinfo {year} {2025})}\BibitemShut {NoStop}%
\bibitem [{\citenamefont {Margalit}\ \emph {et~al.}(2019)\citenamefont
  {Margalit}, \citenamefont {Zhou}, \citenamefont {Machluf}, \citenamefont
  {Japha}, \citenamefont {Moukouri},\ and\ \citenamefont
  {Folman}}]{margalit_analysis_2019}%
  \BibitemOpen
  \bibfield  {author} {\bibinfo {author} {\bibfnamefont {Y.}~\bibnamefont
  {Margalit}}, \bibinfo {author} {\bibfnamefont {Z.}~\bibnamefont {Zhou}},
  \bibinfo {author} {\bibfnamefont {S.}~\bibnamefont {Machluf}}, \bibinfo
  {author} {\bibfnamefont {Y.}~\bibnamefont {Japha}}, \bibinfo {author}
  {\bibfnamefont {S.}~\bibnamefont {Moukouri}}, \ and\ \bibinfo {author}
  {\bibfnamefont {R.}~\bibnamefont {Folman}},\ }\href {\doibase
  10.1088/1367-2630/ab2fdc} {\bibfield  {journal} {\bibinfo  {journal} {New
  Journal of Physics}\ }\textbf {\bibinfo {volume} {21}},\ \bibinfo {pages}
  {073040} (\bibinfo {year} {2019})}\BibitemShut {NoStop}%
\bibitem [{\citenamefont {Zimmermann}\ \emph {et~al.}(2018)\citenamefont
  {Zimmermann}, \citenamefont {Efremov}, \citenamefont {Roura}, \citenamefont
  {Schleich}, \citenamefont {DeSavage}, \citenamefont {Davis}, \citenamefont
  {Srinivasan}, \citenamefont {Narducci}, \citenamefont {Werner},\ and\
  \citenamefont {Rasel}}]{zimmermann_t3_interferometer_2018}%
  \BibitemOpen
  \bibfield  {author} {\bibinfo {author} {\bibfnamefont {M.}~\bibnamefont
  {Zimmermann}}, \bibinfo {author} {\bibfnamefont {M.~A.}\ \bibnamefont
  {Efremov}}, \bibinfo {author} {\bibfnamefont {A.}~\bibnamefont {Roura}},
  \bibinfo {author} {\bibfnamefont {W.~P.}\ \bibnamefont {Schleich}}, \bibinfo
  {author} {\bibfnamefont {S.~A.}\ \bibnamefont {DeSavage}}, \bibinfo {author}
  {\bibfnamefont {J.~P.}\ \bibnamefont {Davis}}, \bibinfo {author}
  {\bibfnamefont {A.}~\bibnamefont {Srinivasan}}, \bibinfo {author}
  {\bibfnamefont {F.~A.}\ \bibnamefont {Narducci}}, \bibinfo {author}
  {\bibfnamefont {S.~A.}\ \bibnamefont {Werner}}, \ and\ \bibinfo {author}
  {\bibfnamefont {E.~M.}\ \bibnamefont {Rasel}},\ }in\ \href {\doibase
  10.1007/978-3-319-64346-5_27} {\emph {\bibinfo {booktitle} {Exploring the
  {World} with the {Laser}: {Dedicated} to {Theodor} {H\"{a}nsch} on his 75th
  birthday}}},\ \bibinfo {editor} {edited by\ \bibinfo {editor} {\bibfnamefont
  {D.}~\bibnamefont {Meschede}}, \bibinfo {editor} {\bibfnamefont
  {T.}~\bibnamefont {Udem}}, \ and\ \bibinfo {editor} {\bibfnamefont
  {T.}~\bibnamefont {Esslinger}}}\ (\bibinfo  {publisher} {Springer
  International Publishing},\ \bibinfo {address} {Cham},\ \bibinfo {year}
  {2018})\ pp.\ \bibinfo {pages} {457--489}\BibitemShut {NoStop}%
\bibitem [{\citenamefont {Paraniak}\ and\ \citenamefont
  {Englert}(2021)}]{shortdoi:gsmsgh}%
  \BibitemOpen
  \bibfield  {author} {\bibinfo {author} {\bibfnamefont {M.~M.}\ \bibnamefont
  {Paraniak}}\ and\ \bibinfo {author} {\bibfnamefont {B.-G.}\ \bibnamefont
  {Englert}},\ }\href {\doibase 10.3390/sym13091660} {\bibfield  {journal}
  {\bibinfo  {journal} {Symmetry}\ }\textbf {\bibinfo {volume} {13}},\ \bibinfo
  {pages} {1660} (\bibinfo {year} {2021})}\BibitemShut {NoStop}%
\bibitem [{\citenamefont {Zhou}\ \emph
  {et~al.}(2025{\natexlab{c}})\citenamefont {Zhou}, \citenamefont {Xiang},\
  and\ \citenamefont {Mazumdar}}]{shortdoi:g9rwcz}%
  \BibitemOpen
  \bibfield  {author} {\bibinfo {author} {\bibfnamefont {R.}~\bibnamefont
  {Zhou}}, \bibinfo {author} {\bibfnamefont {Q.}~\bibnamefont {Xiang}}, \ and\
  \bibinfo {author} {\bibfnamefont {A.}~\bibnamefont {Mazumdar}},\ }\href
  {\doibase 10.1103/PhysRevA.111.052207} {\bibfield  {journal} {\bibinfo
  {journal} {Physical Review A}\ }\textbf {\bibinfo {volume} {111}},\ \bibinfo
  {pages} {052207} (\bibinfo {year} {2025}{\natexlab{c}})}\BibitemShut
  {NoStop}%
\bibitem [{\citenamefont {Hsu}\ \emph {et~al.}(2011)\citenamefont {Hsu},
  \citenamefont {Berrondo},\ and\ \citenamefont {Van~Huele}}]{shortdoi:d9pb92}%
  \BibitemOpen
  \bibfield  {author} {\bibinfo {author} {\bibfnamefont {B.~C.}\ \bibnamefont
  {Hsu}}, \bibinfo {author} {\bibfnamefont {M.}~\bibnamefont {Berrondo}}, \
  and\ \bibinfo {author} {\bibfnamefont {J.-F.~S.}\ \bibnamefont {Van~Huele}},\
  }\href {\doibase 10.1103/PhysRevA.83.012109} {\bibfield  {journal} {\bibinfo
  {journal} {Physical Review A}\ }\textbf {\bibinfo {volume} {83}},\ \bibinfo
  {pages} {012109} (\bibinfo {year} {2011})}\BibitemShut {NoStop}%
\bibitem [{\citenamefont {Palmer}\ and\ \citenamefont
  {Hogan}(2019{\natexlab{a}})}]{shortdoi:gf432r}%
  \BibitemOpen
  \bibfield  {author} {\bibinfo {author} {\bibfnamefont {J.~E.}\ \bibnamefont
  {Palmer}}\ and\ \bibinfo {author} {\bibfnamefont {S.~D.}\ \bibnamefont
  {Hogan}},\ }\href {\doibase 10.1080/00268976.2019.1607916} {\bibfield
  {journal} {\bibinfo  {journal} {Molecular Physics}\ }\textbf {\bibinfo
  {volume} {117}},\ \bibinfo {pages} {3108} (\bibinfo {year}
  {2019}{\natexlab{a}})}\BibitemShut {NoStop}%
\bibitem [{\citenamefont {Palmer}\ and\ \citenamefont
  {Hogan}(2019{\natexlab{b}})}]{shortdoi:gg2bbk}%
  \BibitemOpen
  \bibfield  {author} {\bibinfo {author} {\bibfnamefont {J.~E.}\ \bibnamefont
  {Palmer}}\ and\ \bibinfo {author} {\bibfnamefont {S.~D.}\ \bibnamefont
  {Hogan}},\ }\href {\doibase 10.1103/PhysRevLett.122.250404} {\bibfield
  {journal} {\bibinfo  {journal} {Physical Review Letters}\ }\textbf {\bibinfo
  {volume} {122}},\ \bibinfo {pages} {250404} (\bibinfo {year}
  {2019}{\natexlab{b}})}\BibitemShut {NoStop}%
\bibitem [{\citenamefont {Tommey}\ and\ \citenamefont
  {Hogan}(2021)}]{shortdoi:gmsf2r}%
  \BibitemOpen
  \bibfield  {author} {\bibinfo {author} {\bibfnamefont {J.~D.~R.}\
  \bibnamefont {Tommey}}\ and\ \bibinfo {author} {\bibfnamefont {S.~D.}\
  \bibnamefont {Hogan}},\ }\href {\doibase 10.1103/PhysRevA.104.033305}
  {\bibfield  {journal} {\bibinfo  {journal} {Physical Review A}\ }\textbf
  {\bibinfo {volume} {104}},\ \bibinfo {pages} {033305} (\bibinfo {year}
  {2021})}\BibitemShut {NoStop}%
\bibitem [{\citenamefont {Chan}\ and\ \citenamefont
  {Martin}(2024)}]{shortdoi:gtddkw}%
  \BibitemOpen
  \bibfield  {author} {\bibinfo {author} {\bibfnamefont {D.~Z.}\ \bibnamefont
  {Chan}}\ and\ \bibinfo {author} {\bibfnamefont {J.~D.~D.}\ \bibnamefont
  {Martin}},\ }\href {\doibase 10.1103/PhysRevA.109.017301} {\bibfield
  {journal} {\bibinfo  {journal} {Physical Review A}\ }\textbf {\bibinfo
  {volume} {109}},\ \bibinfo {pages} {017301} (\bibinfo {year}
  {2024})}\BibitemShut {NoStop}%
\bibitem [{\citenamefont {Monroe}\ \emph {et~al.}(1990)\citenamefont {Monroe},
  \citenamefont {Swann}, \citenamefont {Robinson},\ and\ \citenamefont
  {Wieman}}]{shortdoi:fjh9zv}%
  \BibitemOpen
  \bibfield  {author} {\bibinfo {author} {\bibfnamefont {C.}~\bibnamefont
  {Monroe}}, \bibinfo {author} {\bibfnamefont {W.}~\bibnamefont {Swann}},
  \bibinfo {author} {\bibfnamefont {H.}~\bibnamefont {Robinson}}, \ and\
  \bibinfo {author} {\bibfnamefont {C.}~\bibnamefont {Wieman}},\ }\href
  {\doibase 10.1103/PhysRevLett.65.1571} {\bibfield  {journal} {\bibinfo
  {journal} {Physical Review Letters}\ }\textbf {\bibinfo {volume} {65}},\
  \bibinfo {pages} {1571} (\bibinfo {year} {1990})}\BibitemShut {NoStop}%
\bibitem [{\citenamefont {Born}\ and\ \citenamefont
  {Wolf}(2019)}]{isbn:9781108477437}%
  \BibitemOpen
  \bibfield  {author} {\bibinfo {author} {\bibfnamefont {M.}~\bibnamefont
  {Born}}\ and\ \bibinfo {author} {\bibfnamefont {E.}~\bibnamefont {Wolf}},\
  }\href@noop {} {\emph {\bibinfo {title} {Principles of Optics:
  Electromagnetic Theory of Propagation, Interference, and Diffraction of
  Light}}},\ \bibinfo {edition} {seventh (expanded) anniversary edition, 60th
  anniversary edition}\ ed.\ (\bibinfo  {publisher} {Cambridge University
  Press},\ \bibinfo {address} {Cambridge},\ \bibinfo {year} {2019})\BibitemShut
  {NoStop}%
\bibitem [{\citenamefont {Gallagher}(1994)}]{isbn:9780521385312}%
  \BibitemOpen
  \bibfield  {author} {\bibinfo {author} {\bibfnamefont {T.~F.}\ \bibnamefont
  {Gallagher}},\ }\href@noop {} {\emph {\bibinfo {title} {Rydberg Atoms}}},\
  \bibinfo {series} {Cambridge Monographs on Atomic, Molecular, and Chemical
  Physics}\ No.~\bibinfo {number} {3}\ (\bibinfo  {publisher} {Cambridge
  University Press},\ \bibinfo {address} {Cambridge ; New York},\ \bibinfo
  {year} {1994})\BibitemShut {NoStop}%
\bibitem [{\citenamefont {Carter}\ and\ \citenamefont
  {Martin}(2013)}]{shortdoi:ggz44h}%
  \BibitemOpen
  \bibfield  {author} {\bibinfo {author} {\bibfnamefont {J.~D.}\ \bibnamefont
  {Carter}}\ and\ \bibinfo {author} {\bibfnamefont {J.~D.~D.}\ \bibnamefont
  {Martin}},\ }\href {\doibase 10.1103/PhysRevA.88.043429} {\bibfield
  {journal} {\bibinfo  {journal} {Physical Review A}\ }\textbf {\bibinfo
  {volume} {88}},\ \bibinfo {pages} {043429} (\bibinfo {year}
  {2013})}\BibitemShut {NoStop}%
\bibitem [{\citenamefont {Holmgren}\ \emph {et~al.}(2010)\citenamefont
  {Holmgren}, \citenamefont {Revelle}, \citenamefont {Lonij},\ and\
  \citenamefont {Cronin}}]{shortdoi:ccsc8n}%
  \BibitemOpen
  \bibfield  {author} {\bibinfo {author} {\bibfnamefont {W.~F.}\ \bibnamefont
  {Holmgren}}, \bibinfo {author} {\bibfnamefont {M.~C.}\ \bibnamefont
  {Revelle}}, \bibinfo {author} {\bibfnamefont {V.~P.~A.}\ \bibnamefont
  {Lonij}}, \ and\ \bibinfo {author} {\bibfnamefont {A.~D.}\ \bibnamefont
  {Cronin}},\ }\href {\doibase 10.1103/PhysRevA.81.053607} {\bibfield
  {journal} {\bibinfo  {journal} {Physical Review A}\ }\textbf {\bibinfo
  {volume} {81}},\ \bibinfo {pages} {053607} (\bibinfo {year}
  {2010})}\BibitemShut {NoStop}%
\bibitem [{\citenamefont {Zimmerman}\ \emph {et~al.}(1979)\citenamefont
  {Zimmerman}, \citenamefont {Littman}, \citenamefont {Kash},\ and\
  \citenamefont {Kleppner}}]{shortdoi:dfv4zd}%
  \BibitemOpen
  \bibfield  {author} {\bibinfo {author} {\bibfnamefont {M.~L.}\ \bibnamefont
  {Zimmerman}}, \bibinfo {author} {\bibfnamefont {M.~G.}\ \bibnamefont
  {Littman}}, \bibinfo {author} {\bibfnamefont {M.~M.}\ \bibnamefont {Kash}}, \
  and\ \bibinfo {author} {\bibfnamefont {D.}~\bibnamefont {Kleppner}},\ }\href
  {\doibase 10.1103/PhysRevA.20.2251} {\bibfield  {journal} {\bibinfo
  {journal} {Physical Review A}\ }\textbf {\bibinfo {volume} {20}},\ \bibinfo
  {pages} {2251} (\bibinfo {year} {1979})}\BibitemShut {NoStop}%
\bibitem [{\citenamefont {Griffiths}(2023)}]{isbn:9781009397759}%
  \BibitemOpen
  \bibfield  {author} {\bibinfo {author} {\bibfnamefont {D.~J.}\ \bibnamefont
  {Griffiths}},\ }\href@noop {} {\emph {\bibinfo {title} {Introduction to
  Electrodynamics}}},\ \bibinfo {edition} {fifth edition}\ ed.\ (\bibinfo
  {publisher} {Cambridge University Press},\ \bibinfo {address} {New York},\
  \bibinfo {year} {2023})\BibitemShut {NoStop}%
\bibitem [{\citenamefont {Wallace}\ \emph {et~al.}(1994)\citenamefont
  {Wallace}, \citenamefont {Dinneen}, \citenamefont {Tan}, \citenamefont
  {Kumarakrishnan}, \citenamefont {Gould},\ and\ \citenamefont
  {Javanainen}}]{shortdoi:bgpnv7}%
  \BibitemOpen
  \bibfield  {author} {\bibinfo {author} {\bibfnamefont {C.~D.}\ \bibnamefont
  {Wallace}}, \bibinfo {author} {\bibfnamefont {T.~P.}\ \bibnamefont
  {Dinneen}}, \bibinfo {author} {\bibfnamefont {K.~Y.~N.}\ \bibnamefont {Tan}},
  \bibinfo {author} {\bibfnamefont {A.}~\bibnamefont {Kumarakrishnan}},
  \bibinfo {author} {\bibfnamefont {P.~L.}\ \bibnamefont {Gould}}, \ and\
  \bibinfo {author} {\bibfnamefont {J.}~\bibnamefont {Javanainen}},\ }\href
  {\doibase 10.1364/JOSAB.11.000703} {\bibfield  {journal} {\bibinfo  {journal}
  {JOSA B}\ }\textbf {\bibinfo {volume} {11}},\ \bibinfo {pages} {703}
  (\bibinfo {year} {1994})}\BibitemShut {NoStop}%
\bibitem [{\citenamefont {Sakurai}\ and\ \citenamefont
  {Napolitano}(2021)}]{isbn:9781108473224}%
  \BibitemOpen
  \bibfield  {author} {\bibinfo {author} {\bibfnamefont {J.~J.}\ \bibnamefont
  {Sakurai}}\ and\ \bibinfo {author} {\bibfnamefont {J.}~\bibnamefont
  {Napolitano}},\ }\href@noop {} {\emph {\bibinfo {title} {Modern {{Quantum
  Mechanics}}}}},\ \bibinfo {edition} {3rd}\ ed.\ (\bibinfo  {publisher}
  {Cambridge University Press},\ \bibinfo {address} {Cambridge},\ \bibinfo
  {year} {2021})\BibitemShut {NoStop}%
\bibitem [{\citenamefont {Feynman}(1998)}]{isbn:9780201360769}%
  \BibitemOpen
  \bibfield  {author} {\bibinfo {author} {\bibfnamefont {R.~P.}\ \bibnamefont
  {Feynman}},\ }\href@noop {} {\emph {\bibinfo {title} {Statistical Mechanics:
  A Set of Lectures}}},\ Advanced Book Classics\ (\bibinfo  {publisher}
  {Westview Press},\ \bibinfo {address} {Boulder, Colo},\ \bibinfo {year}
  {1998})\BibitemShut {NoStop}%
\bibitem [{\citenamefont {Roura}\ \emph {et~al.}(2014)\citenamefont {Roura},
  \citenamefont {Zeller},\ and\ \citenamefont {Schleich}}]{shortdoi:gtrpxw}%
  \BibitemOpen
  \bibfield  {author} {\bibinfo {author} {\bibfnamefont {A.}~\bibnamefont
  {Roura}}, \bibinfo {author} {\bibfnamefont {W.}~\bibnamefont {Zeller}}, \
  and\ \bibinfo {author} {\bibfnamefont {W.~P.}\ \bibnamefont {Schleich}},\
  }\href {\doibase 10.1088/1367-2630/16/12/123012} {\bibfield  {journal}
  {\bibinfo  {journal} {New Journal of Physics}\ }\textbf {\bibinfo {volume}
  {16}},\ \bibinfo {pages} {123012} (\bibinfo {year} {2014})}\BibitemShut
  {NoStop}%
\bibitem [{\citenamefont {Schwinger}\ \emph {et~al.}(1988)\citenamefont
  {Schwinger}, \citenamefont {Scully},\ and\ \citenamefont
  {Englert}}]{shortdoi:cdxkgn}%
  \BibitemOpen
  \bibfield  {author} {\bibinfo {author} {\bibfnamefont {J.}~\bibnamefont
  {Schwinger}}, \bibinfo {author} {\bibfnamefont {M.~O.}\ \bibnamefont
  {Scully}}, \ and\ \bibinfo {author} {\bibfnamefont {B.~G.}\ \bibnamefont
  {Englert}},\ }\href {\doibase 10.1007/BF01384847} {\bibfield  {journal}
  {\bibinfo  {journal} {Zeitschrift f{\"u}r Physik D Atoms, Molecules and
  Clusters}\ }\textbf {\bibinfo {volume} {10}},\ \bibinfo {pages} {135}
  (\bibinfo {year} {1988})}\BibitemShut {NoStop}%
\bibitem [{\citenamefont {Zimmermann}\ \emph {et~al.}(2019)\citenamefont
  {Zimmermann}, \citenamefont {Efremov}, \citenamefont {Zeller}, \citenamefont
  {Schleich}, \citenamefont {Davis},\ and\ \citenamefont
  {Narducci}}]{shortdoi:gpcbfg}%
  \BibitemOpen
  \bibfield  {author} {\bibinfo {author} {\bibfnamefont {M.}~\bibnamefont
  {Zimmermann}}, \bibinfo {author} {\bibfnamefont {M.~A.}\ \bibnamefont
  {Efremov}}, \bibinfo {author} {\bibfnamefont {W.}~\bibnamefont {Zeller}},
  \bibinfo {author} {\bibfnamefont {W.~P.}\ \bibnamefont {Schleich}}, \bibinfo
  {author} {\bibfnamefont {J.~P.}\ \bibnamefont {Davis}}, \ and\ \bibinfo
  {author} {\bibfnamefont {F.~A.}\ \bibnamefont {Narducci}},\ }\href {\doibase
  10.1088/1367-2630/ab2e8c} {\bibfield  {journal} {\bibinfo  {journal} {New
  Journal of Physics}\ }\textbf {\bibinfo {volume} {21}},\ \bibinfo {pages}
  {073031} (\bibinfo {year} {2019})}\BibitemShut {NoStop}%
\bibitem [{git()}]{github:our_repo}%
  \BibitemOpen
  \href@noop {} {}\bibinfo {note} {See code at
  \url{https://github.com/jddmartin/humpty_is_3d_supp/}}\BibitemShut {NoStop}%
\bibitem [{\citenamefont {{Zinn-Justin}}(2011)}]{isbn:9780198566755}%
  \BibitemOpen
  \bibfield  {author} {\bibinfo {author} {\bibfnamefont {J.}~\bibnamefont
  {{Zinn-Justin}}},\ }\href@noop {} {\emph {\bibinfo {title} {Path Integrals in
  Quantum Mechanics}}},\ \bibinfo {edition} {repr}\ ed.,\ Oxford Graduate
  Texts\ (\bibinfo  {publisher} {Univ. Press},\ \bibinfo {address} {Oxford},\
  \bibinfo {year} {2011})\BibitemShut {NoStop}%
\bibitem [{\citenamefont {Itano}\ \emph {et~al.}(1993)\citenamefont {Itano},
  \citenamefont {Bergquist}, \citenamefont {Bollinger}, \citenamefont
  {Gilligan}, \citenamefont {Heinzen}, \citenamefont {Moore}, \citenamefont
  {Raizen},\ and\ \citenamefont {Wineland}}]{shortdoi:c9q26q}%
  \BibitemOpen
  \bibfield  {author} {\bibinfo {author} {\bibfnamefont {W.~M.}\ \bibnamefont
  {Itano}}, \bibinfo {author} {\bibfnamefont {J.~C.}\ \bibnamefont
  {Bergquist}}, \bibinfo {author} {\bibfnamefont {J.~J.}\ \bibnamefont
  {Bollinger}}, \bibinfo {author} {\bibfnamefont {J.~M.}\ \bibnamefont
  {Gilligan}}, \bibinfo {author} {\bibfnamefont {D.~J.}\ \bibnamefont
  {Heinzen}}, \bibinfo {author} {\bibfnamefont {F.~L.}\ \bibnamefont {Moore}},
  \bibinfo {author} {\bibfnamefont {M.~G.}\ \bibnamefont {Raizen}}, \ and\
  \bibinfo {author} {\bibfnamefont {D.~J.}\ \bibnamefont {Wineland}},\ }\href
  {\doibase 10.1103/PhysRevA.47.3554} {\bibfield  {journal} {\bibinfo
  {journal} {Physical Review A}\ }\textbf {\bibinfo {volume} {47}},\ \bibinfo
  {pages} {3554} (\bibinfo {year} {1993})}\BibitemShut {NoStop}%
\bibitem [{\citenamefont {Glick}\ and\ \citenamefont
  {Kovachy}(2024)}]{shortdoi:gt63nj}%
  \BibitemOpen
  \bibfield  {author} {\bibinfo {author} {\bibfnamefont {J.}~\bibnamefont
  {Glick}}\ and\ \bibinfo {author} {\bibfnamefont {T.}~\bibnamefont
  {Kovachy}},\ }\href {\doibase 10.48550/arXiv.2407.11446} {\enquote {\bibinfo
  {title} {Feynman {{Diagrams}} for {{Matter Wave Interferometry}}},}\ }
  (\bibinfo {year} {2024}),\ \Eprint {http://arxiv.org/abs/2407.11446}
  {arXiv:2407.11446} \BibitemShut {NoStop}%
\bibitem [{\citenamefont {{Barrag{\'a}n-Gil}}\ and\ \citenamefont
  {Walser}(2018)}]{shortdoi:g9wqxf}%
  \BibitemOpen
  \bibfield  {author} {\bibinfo {author} {\bibfnamefont {L.~F.}\ \bibnamefont
  {{Barrag{\'a}n-Gil}}}\ and\ \bibinfo {author} {\bibfnamefont
  {R.}~\bibnamefont {Walser}},\ }\href {\doibase 10.1119/1.5008268} {\bibfield
  {journal} {\bibinfo  {journal} {American Journal of Physics}\ }\textbf
  {\bibinfo {volume} {86}},\ \bibinfo {pages} {22} (\bibinfo {year}
  {2018})}\BibitemShut {NoStop}%
\bibitem [{\citenamefont {Gallagher}\ and\ \citenamefont
  {Cooke}(1979)}]{shortdoi:fctf3r}%
  \BibitemOpen
  \bibfield  {author} {\bibinfo {author} {\bibfnamefont {T.~F.}\ \bibnamefont
  {Gallagher}}\ and\ \bibinfo {author} {\bibfnamefont {W.~E.}\ \bibnamefont
  {Cooke}},\ }\href {\doibase 10.1103/PhysRevLett.42.835} {\bibfield  {journal}
  {\bibinfo  {journal} {Physical Review Letters}\ }\textbf {\bibinfo {volume}
  {42}},\ \bibinfo {pages} {835} (\bibinfo {year} {1979})}\BibitemShut
  {NoStop}%
\bibitem [{\citenamefont {Branden}\ \emph {et~al.}(2009)\citenamefont
  {Branden}, \citenamefont {Juhasz}, \citenamefont {Mahlokozera}, \citenamefont
  {Vesa}, \citenamefont {Wilson}, \citenamefont {Zheng}, \citenamefont
  {Kortyna},\ and\ \citenamefont {Tate}}]{shortdoi:b92t65}%
  \BibitemOpen
  \bibfield  {author} {\bibinfo {author} {\bibfnamefont {D.~B.}\ \bibnamefont
  {Branden}}, \bibinfo {author} {\bibfnamefont {T.}~\bibnamefont {Juhasz}},
  \bibinfo {author} {\bibfnamefont {T.}~\bibnamefont {Mahlokozera}}, \bibinfo
  {author} {\bibfnamefont {C.}~\bibnamefont {Vesa}}, \bibinfo {author}
  {\bibfnamefont {R.~O.}\ \bibnamefont {Wilson}}, \bibinfo {author}
  {\bibfnamefont {M.}~\bibnamefont {Zheng}}, \bibinfo {author} {\bibfnamefont
  {A.}~\bibnamefont {Kortyna}}, \ and\ \bibinfo {author} {\bibfnamefont
  {D.~A.}\ \bibnamefont {Tate}},\ }\href {\doibase
  10.1088/0953-4075/43/1/015002} {\bibfield  {journal} {\bibinfo  {journal}
  {Journal of Physics B: Atomic, Molecular and Optical Physics}\ }\textbf
  {\bibinfo {volume} {43}},\ \bibinfo {pages} {015002} (\bibinfo {year}
  {2009})}\BibitemShut {NoStop}%
\bibitem [{\citenamefont {Thoumany}\ \emph {et~al.}(2009)\citenamefont
  {Thoumany}, \citenamefont {Germann}, \citenamefont {H{\"a}nsch},
  \citenamefont {Stania}, \citenamefont {Urbonas},\ and\ \citenamefont
  {Becker}}]{shortdoi:d5f2xr}%
  \BibitemOpen
  \bibfield  {author} {\bibinfo {author} {\bibfnamefont {P.}~\bibnamefont
  {Thoumany}}, \bibinfo {author} {\bibfnamefont {{\relax Th}.}~\bibnamefont
  {Germann}}, \bibinfo {author} {\bibfnamefont {T.}~\bibnamefont {H{\"a}nsch}},
  \bibinfo {author} {\bibfnamefont {G.}~\bibnamefont {Stania}}, \bibinfo
  {author} {\bibfnamefont {L.}~\bibnamefont {Urbonas}}, \ and\ \bibinfo
  {author} {\bibfnamefont {{\relax Th}.}~\bibnamefont {Becker}},\ }\href
  {\doibase 10.1080/09500340903180525} {\bibfield  {journal} {\bibinfo
  {journal} {Journal of Modern Optics}\ }\textbf {\bibinfo {volume} {56}},\
  \bibinfo {pages} {2055} (\bibinfo {year} {2009})}\BibitemShut {NoStop}%
\bibitem [{\citenamefont {Johnson}\ \emph {et~al.}(2010)\citenamefont
  {Johnson}, \citenamefont {Majeed}, \citenamefont {Sanguinetti}, \citenamefont
  {Becker},\ and\ \citenamefont {Varcoe}}]{shortdoi:dh5fxf}%
  \BibitemOpen
  \bibfield  {author} {\bibinfo {author} {\bibfnamefont {L.~A.~M.}\
  \bibnamefont {Johnson}}, \bibinfo {author} {\bibfnamefont {H.~O.}\
  \bibnamefont {Majeed}}, \bibinfo {author} {\bibfnamefont {B.}~\bibnamefont
  {Sanguinetti}}, \bibinfo {author} {\bibfnamefont {T.}~\bibnamefont {Becker}},
  \ and\ \bibinfo {author} {\bibfnamefont {B.~T.~H.}\ \bibnamefont {Varcoe}},\
  }\href {\doibase 10.1088/1367-2630/12/6/063028} {\bibfield  {journal}
  {\bibinfo  {journal} {New Journal of Physics}\ }\textbf {\bibinfo {volume}
  {12}},\ \bibinfo {pages} {063028} (\bibinfo {year} {2010})}\BibitemShut
  {NoStop}%
\bibitem [{\citenamefont {Li}\ \emph {et~al.}(2003)\citenamefont {Li},
  \citenamefont {Mourachko}, \citenamefont {Noel},\ and\ \citenamefont
  {Gallagher}}]{shortdoi:dntc3s}%
  \BibitemOpen
  \bibfield  {author} {\bibinfo {author} {\bibfnamefont {W.}~\bibnamefont
  {Li}}, \bibinfo {author} {\bibfnamefont {I.}~\bibnamefont {Mourachko}},
  \bibinfo {author} {\bibfnamefont {M.~W.}\ \bibnamefont {Noel}}, \ and\
  \bibinfo {author} {\bibfnamefont {T.~F.}\ \bibnamefont {Gallagher}},\ }\href
  {\doibase 10.1103/PhysRevA.67.052502} {\bibfield  {journal} {\bibinfo
  {journal} {Physical Review A}\ }\textbf {\bibinfo {volume} {67}},\ \bibinfo
  {pages} {052502} (\bibinfo {year} {2003})}\BibitemShut {NoStop}%
\bibitem [{\citenamefont {Han}\ \emph {et~al.}(2006)\citenamefont {Han},
  \citenamefont {Jamil}, \citenamefont {Norum}, \citenamefont {Tanner},\ and\
  \citenamefont {Gallagher}}]{shortdoi:dc3462}%
  \BibitemOpen
  \bibfield  {author} {\bibinfo {author} {\bibfnamefont {J.}~\bibnamefont
  {Han}}, \bibinfo {author} {\bibfnamefont {Y.}~\bibnamefont {Jamil}}, \bibinfo
  {author} {\bibfnamefont {D.~V.~L.}\ \bibnamefont {Norum}}, \bibinfo {author}
  {\bibfnamefont {P.~J.}\ \bibnamefont {Tanner}}, \ and\ \bibinfo {author}
  {\bibfnamefont {T.~F.}\ \bibnamefont {Gallagher}},\ }\href {\doibase
  10.1103/PhysRevA.74.054502} {\bibfield  {journal} {\bibinfo  {journal}
  {Physical Review A}\ }\textbf {\bibinfo {volume} {74}},\ \bibinfo {pages}
  {054502} (\bibinfo {year} {2006})}\BibitemShut {NoStop}%
\bibitem [{\citenamefont {Robinett}(1996)}]{shortdoi:cd9r2k}%
  \BibitemOpen
  \bibfield  {author} {\bibinfo {author} {\bibfnamefont {R.~W.}\ \bibnamefont
  {Robinett}},\ }\href {\doibase 10.1119/1.18179} {\bibfield  {journal}
  {\bibinfo  {journal} {American Journal of Physics}\ }\textbf {\bibinfo
  {volume} {64}},\ \bibinfo {pages} {803} (\bibinfo {year} {1996})}\BibitemShut
  {NoStop}%
\bibitem [{\citenamefont {Barton}(1986)}]{shortdoi:dnrjcn}%
  \BibitemOpen
  \bibfield  {author} {\bibinfo {author} {\bibfnamefont {G.}~\bibnamefont
  {Barton}},\ }\href {\doibase 10.1016/0003-4916(86)90142-9} {\bibfield
  {journal} {\bibinfo  {journal} {Annals of Physics}\ }\textbf {\bibinfo
  {volume} {166}},\ \bibinfo {pages} {322} (\bibinfo {year}
  {1986})}\BibitemShut {NoStop}%
\bibitem [{\citenamefont {Pauli}(2000)}]{isbn:9780486414621}%
  \BibitemOpen
  \bibfield  {author} {\bibinfo {author} {\bibfnamefont {W.}~\bibnamefont
  {Pauli}},\ }\href@noop {} {\emph {\bibinfo {title} {Wave Mechanics}}},\
  \bibinfo {series} {Pauli Lectures on Physics}\ No.\ \bibinfo {number} {v. 5}\
  (\bibinfo  {publisher} {Dover Publications},\ \bibinfo {address} {Mineola,
  N.Y},\ \bibinfo {year} {2000})\BibitemShut {NoStop}%
\end{thebibliography}
\end{document}